\documentclass{aa}  

\def\oiii{[O\,{\sc iii}]}
\def\oii{[O\,{\sc ii}]}
\def\oi{[O\,{\sc i}]}

\def\ha{H$\alpha$}

\def\nii{[N\,{\sc ii}]}
\def\sii{[S\,{\sc ii}]}

\def\kms{\relax \ifmmode {\,\rm km\,s}^{-1}\else \,km\,s$^{-1}$\fi}
\def\deg{$^\circ$}
\usepackage{graphicx}
\usepackage{txfonts}
\usepackage{natbib}
\bibpunct{(}{)}{;}{a}{}{,}
\begin{document} 

\title{New insights into the outflows from R Aquarii}

  %\subtitle{}

    \author{T. Liimets
          \inst{1,2}
          \and
          R.~L.~M. Corradi
          \inst{3,4}
          \and
          D. Jones
          \inst{4,5}
          \and
           K. Verro
          \inst{1,6}
          \and
           M. Santander-Garc{\'{\i}}a
          \inst{7}         
           \and
          I. Kolka
          \inst{1}         
                \and
         M. Sidonio
          \inst{8}
          \and
         E. Kankare
          \inst{9}
          \and
         J. Kankare
          \inst{9}
          \and
          T. Pursimo
          \inst{10}
                     \and
         P.~A. Wilson
          \inst{11,12}
           }

\institute{Tartu Observatory, T\~oravere, 61602, Estonia\\
              \email{tiina@obs.ee}
         \and
              Institute of Physics, University of Tartu, Ravila 14c, 50411, Estonia     %\email{tiina@obs.ee}
         \and 
         GRANTECAN, Cuesta de San Jos\'e s/n, E-38712, Bre\~na Baja, La Palma, Spain
         %\email{romano.corradi@gtc.iac.es}
         \and     
              Instituto de Astrof{\'{\i}}sica de Canarias, E-38200 La Laguna, Tenerife, Spain
          \and 
              Departamento de Astrof{\'{\i}}sica, Universidad de La Laguna, E-38206 La Laguna, Tenerife, Spain % \email{}          	    
         \and Kapteyn Instituut, Rijksuniversiteit Groningen, Landleven 12, 9747AD Groningen, The Netherlands
          \and 
          Observatorio Astron\'omico Nacional (OAN-IGN), C/ Alfonso XII, 3, 28014, Madrid, Spain
          \and
          Terroux Observatory, Canberra, Australia
 	\and
	Astrophysics Research Centre, School of Mathematics and Physics, Queen's University Belfast, Belfast BT7 1NN, UK
          \and 
              Nordic Optical Telescope, Apartado 474, ES-38700 Santa Cruz de La Palma, Spain
	\and
	Leiden Observatory, Leiden University, Postbus 9513, 2300 RA Leiden, The Netherlands
	\and
	CNRS, UMR 7095, Institut d'Astrophysique de Paris, $98^{\mathrm{bis}}$ Boulevard Arago, F-75014 Paris, France
             }

   \date{Received ; accepted }

% \abstract{}{}{}{}{} 
% 5 {} token are mandatory
 
  \abstract
  % context heading (optional), {} leave it empty if necessary  
   {R Aquarii is a symbiotic binary surrounded by a large and complex nebula with a prominent curved jet. 
   It is one of the closest known symbiotic systems, and therefore offers a unique opportunity to study the central 
   regions of these systems and the formation and evolution of astrophysical jets.}
   {The evolution of the central jet and outer nebula of R Aqr is studied taking advantage of a long term monitoring campaign of optical imaging, as well as of 
   high-resolution integral field spectroscopy.}
   {Narrow-band images acquired over a period of more than 21 years were compared in order to study the expansion 
   and evolution of all components of the R Aqr nebula. The magnification method is used to derive the kinematic ages of the 
   features that appear to expand radially. 
   %% while strong deviations from a uniform, radial expansion are found for other components. 
   Integral field spectroscopy of the \oiii~5007~\AA{} emission is used to study the velocity structure of the central regions of the jet.}
   {New extended features, further out than the previously known hourglass nebula, are detected. The kinematic distance to 
   R Aqr is calculated to be 178 pc using the expansion of the large hourglass nebula. 
   This nebula of R Aqr is found to be roughly 650 years old, while the inner regions have ages ranging from 125 to 290 years. 
   The outer nebula is found to be well described by a ballistic expansion, while for most components of the 
   jet strong deviations from such behaviour are found.
   We find that the Northern jet is mostly red-shifted while its Southern part is blue-shifted, apparently at odds 
   with findings from previous studies but almost certainly a consequence of the complex nature of the jet and 
   variations in ionisation and illumination between observations.  
   }
   {}


   \maketitle
%
%________________________________________________________________

\section{Introduction}
\label{S-intro}

Symbiotic stars are interacting binaries composed of a hot component, usually a white dwarf, and a mass losing red giant. The large mass loss 
from these evolved companions, the fast winds of the white dwarfs, and the occurrence of nova-like explosions, 
produce a rich circumstellar environment often 
taking the form of bipolar nebulae, collimated jets, or generally complex ejecta.

R Aquarii (R Aqr) is a symbiotic binary system, consisting of a M7\textsc{iii} Mira variable and a white dwarf, 
surrounded by complex nebular structures extending across several arcminutes.  At large scales, 
R Aqr appears as a bipolar, hourglass-like nebula with a prominent toroidal structure 
at its waist, within which a curved jet-like structure is found. At a distance of about 200 pc R Aqr is the closest known symbiotic binary, 
and therefore provides a unique opportunity to study 
in detail the evolution of a stellar outflows.

%% NEBULA INTRO
The hourglass nebula of R Aqr was first discovered by \citet{1922PAAS....4..319L} and  repeated observations have revealed that it is 
expanding - at a first approximation - in a ballistic way 
\citep{1985A&A...148..274S}.  The expansion of the large-scale nebula was used by \citet{1944MWOAR..16....1A} to calculate a 
kinematical age of 600 yr. \citet{1985A&A...148..274S} refine this value to 640 yrs by applying 
a kinematical model using a hourglass geometry, with an equatorial expansion velocity of 55 \kms{}, and assuming  that  the expansion velocity 
at each point of the nebula is proportional to the distance from the centre.  

%% JET INTRO
The presence of the central jet in R Aqr was first remarked upon by \cite{1980PASP...92..275W}, however  
\citet{1999ApJ...514..895H} showed that the jet was present in observations taken as early as 1934. 
Since these earlier observations, the large-scale S-shape of the jet has remained unchanged, 
while at smaller scales its  appearance varies greatly even on short timescales 
\citep[][this work]{1991ApJ...369L..67P,1988AJ.....95.1478M,1990ApJ...351L..17H,1999ApJ...522..297H,2007ApJ...664.1079K}.
A detailed investigation of the innermost $5''$ of the jet has been carried out using 
high resolution radio data (e.g \citet{1989ApJ...346..991K,2004A&A...424..157M}). 
However, it has been demonstrated that at different 
wavelengths the appearance and the radial velocity pattern of the jet varies dramatically 
\citep{1991ApJ...369L..67P,1993ApJ...411..235H,1982ApJ...258L..35S,1985A&A...148..274S,1990ApJ...351L..17H,1999ApJ...522..297H}.

In this article, we present a detailed study of the R Aqr nebula based on deep, narrow-band emission line imaging, 
acquired over a period of more than two decades, as well as on high spectral resolution integral field spectroscopy of the \mbox{\oiii\ 5007 \AA} 
 emission from the central regions.  The data and data reduction is presented in Section~\ref{S-data}, with the image processing 
techniques employed in our analysis in Section~\ref{S-pr}.  Sections~\ref{S-of} and \ref{S-j} contain the results of our analysis of 
the multi-epoch imaging, while Section~\ref{S-rv} is dedicated to the results from the integral field spectroscopy. 
Finally, in Section~\ref{S-conc}, we present our conclusions.

%%%%%%%%%%%%%%%%%%%%%%%%%%%%%%%%%%%%%%%%%%%
%%%%%%%%%%%%%%%%%%%%%%%%%%%%%%%%%%%%%%%%%%%

\section{Observations and data reduction}\label{S-data}

\subsection{Imaging}\label{S-ima}

The imaging data presented in this paper was collected over more than two decades at various observatories.  Most
of the data were obtained with the 2.6m Nordic Optical Telescope (NOT)
using the Andalucia Faint Object Spectrograph and Camera
(ALFOSC). ALFOSC has a pixel scale $0''.19$ pix$^{-1}$ and a field of
view (FOV) of $6'.4\times6'.4$.  Previously, in 1991, we obtained a single
\ha+\nii\ image with European Southern Observatory's (ESO) New Technology Telescope (NTT) equipped
with ESO Multi-Mode Instrument (EMMI, $0''.35$ pix$^{-1}$ and a FOV of
$6'.2\times6'.2$ \citep{1986SPIE..627..339D}). 
More recent data  were obtained in 2012 with ESO's Very Large Telescope (VLT) and its Focal Reducer/low
dispersion Spectrograph 2 (FORS2 \citep{1998Msngr..94....1A}). The Standard Resolution
collimator of FORS2 was used resulting in a pixel scale of $0''.25$ pix$^{-1}$ and a
FOV of $6'.8\times6'.8$.  At all epochs several narrow band filters were
used.  Due to the moderate radial velocities present in the R Aquarii jet and
bipolar nebula, all filters
% see more from opt_filters.numbers
include all the light from corresponding emission lines, except the
2002 \nii\ filter which only includes radial velocities larger than +45~\kms 
considering the \nii\ 6583~\AA\ rest wavelength. Filters centred at \ha\ also include emission
from the \nii$\lambda\lambda$6548, 6583~\AA\ doublet. Details of 
the central wavelengths (CW) and full widths at half maxima
(FWHM) of all filters employed, as well as other observational details can be found in
Table~\ref{T-imaobs}. 
As can be seen from the table, most of the data were taken with the NOT+ALFOSC, 
under good seeing conditions. Airmass mostly never exceeded 1.5. 
This allows safe comparison of this homogenous set of images.
All the individual frames were reduced (bias, flat field correction) using
standard routines in IRAF\footnote{IRAF is distributed by the National 
Optical Astronomy Observatory, which is operated by the Association 
of Universities for Research in Astronomy (AURA) under cooperative 
agreement with the National Science Foundation.}.

\begin{table*}
\caption{Log of the imaging data. The first column lists the start of the
  observing run. In the second column JD is the Julian Date at mid point of the observations.  
Column 3 lists the telescope and instrument used. Column 4 contains the nebular lines included and the central wavelength and the
  FWHM of the filter. Column 5 is the total exposure time, column 6 is
  the number of frames added together.  In column 7 there is the FWHM seeing as measured on the detector.}
\label{T-imaobs}
\centering
\begin{tabular}{llllrrr}
\hline\hline
Date & JD & Telescope+  &Filter                      &  Total exp.   &  Nr. of    &  Seeing   \\ 
         &       &   Instrument   &  CW/FWHM (\AA) &  time (sec)  &  frames  &  ($''$)\\ 
\hline
1991-07-06\tablefootmark{a}  & 2448443.794005 & NTT+EMMI       & \ha+\nii\ 6568/73 & 300, 30, 1  & 1, 1, 1 & 1.1  \\    % filter #596
1998-09-04    &  2451061.541661  & NOT+ALFOSC & \ha+\nii\ 6577/180& 100 & 1 & 1.3      \\     
2009-07-08$^\ast\tablefootmark{b}$    & 2455021.682257  & NOT+ALFOSC & \ha+\nii\ 6577/180& 100, 2 & 1, 1  &  1.3, 1.2 \\    
2012-09-05   & 2456175.775466    & VLT+FORS2   & \ha+\nii\ 6563/61  & 90 & 3 & 0.7\\   \\  
1997-07-13\tablefootmark{a}    & 2450643.774977  & NOT+ALFOSC & \nii\ 6584/10 & 120, 20, 5 & 1, 1, 1 & 1.1    \\
2002-06-26    & 2452452.731273   & NOT+ALFOSC & \nii\ 6588/9 & 600  & 1 &  0.8   \\   \\  % NII-IAC#41
2002-06-27    &  2452453.769479   & NOT+ALFOSC & \oiii\ 5008/30& 60, 30, 10 & 1, 1, 1 & 0.9  \\  % IAC#6-OIII
2007-09-05$^\ast$  & 2454349.620787 & NOT+ALFOSC &  \oiii\ 5007/30& 400, 30 &  1, 1& 0.7, 0.6 \\    % OIII, \oiii\ 501_3, #90
2009-07-08 &2455021.683102 & NOT+ALFOSC& \oiii\  5007/30& 60 &1 & 1.2  \\ 
2009-08-24       & 2455068.647587   &NOT+ALFOSC & \oiii\ 5007/30& 600 & 2 & 0.6 \\   
2011-09-05       & 2455810.571493   &NOT+ALFOSC & \oiii\ 5007/30& 180, 40 & 1, 1 & 0.9, 0.8 \\     
2012-09-05   & 2456175.775466    & VLT+FORS2   &\oiii\ 5001/57 & 300, 1 & 3, 1 & 0.7\\ \\
1997-07-15\tablefootmark{a}    & 2450645.740139 & NOT+ALFOSC & \oii\ 3727/30 & 900 & 3 & 1.1  \\  
1998-09-04\tablefootmark{c}     &  2451061.541661  & NOT+ALFOSC &  \oii\  3727/30 & 300 & 1 & 1.3      \\  %  \oii\ IAC
2002-06-27    &  2452453.769479   & NOT+ALFOSC &  \oii\  3725/50& 900  & 1 & 1.3  \\  % OII-IAC#4
2007-09-05$^\ast$  & 2454349.620787 & NOT+ALFOSC &   \oii\  3726/51& 600 &  1& 0.7 \\    % \oii\ 373_5 # 30
2009-08-24       & 2455068.647587   &NOT+ALFSOC &  \oii\  3726/51 & 1200 & 2 & 0.8 \\  % \oii\ 373_5 # 30
2011-09-05       & 2455810.571493   &NOT+ALFSOC &  \oii\  3726/51 & 600 & 1 & 0.9 \\  % \oii\ 373_5 # 30
2012-09-05   & 2456175.775466    & VLT+FORS2   &  \oii\ 3717/73  & 540, 10 & 3, 1 & 0.8\\  \\  % +44

2007-09-05$^\ast$& 2454349.620787 & NOT+ALFOSC &   \oi\  6308/29& 120, 30 &  1, 1& 0.6 \\   % OI IAC037
2009-08-24        & 2455068.647587   &NOT+ALFSOC &  \oi\  6300/30 & 600 & 2 & 0.6 \\  % \oi\ 630_3' # 87
\hline
\end{tabular}
\tablefoot{
\tablefoottext{a}{Published in \citet{2003ASPC..303..486N} and  \citet{2003ASPC..303..423G}.}\\
\tablefoottext{b}{$^\ast$ indicate the reference epoch for the pixel by pixel matching for a given filter (see Section \ref{S-pr}).}\\
\tablefoottext{c}{Published in \citet{2003ASPC..303..393C}.}
}
\end{table*}

An additional set of deep observations was obtained on October 2, 8, and November 2, 3 2016 in Terroux Observatory, 
Canberra, Australia using a 30cm f3.8 Newtonian telescope with a CCD camera and narrow 
band \oiii\ (5010~\AA, FWHM=60~\AA) and \ha\ (6560~\AA, FWHM=60~\AA) filters. The pixel scale was $0''.84$ pix$^{-1}$. 
34 frames for a total exposure time of 7.6h were obtained in \oiii, while 32 frames adding up to 6.8 hours were taken in \ha. 
Frames were flat field corrected and 
median combined.

\subsection{Spectroscopy}
\label{S-sp}

Integral Field Unit (IFU) spectroscopy was obtained on October 3, 2012, with
the VLT equipped with the Fibre Large Array Multi Element Spectrograph
FLAMES \citep{2002Msngr.110....1P} in GIRAFFE/ARGUS mode. The ARGUS
IFU provides continuous spectral coverage for a $11''.4\times7''.3$
FOV (formed from an array of $0''.52$ lenslets).  The high resolution grating HR08
was used in the spectral range from 4920~\AA\ to 5160~\AA, covering both \oiii\ lines at 5007~\AA\ and 4959~\AA. The spectral
reciprocal dispersion was 0.05~\AA\ pix$^{-1}$. % 0.005 nm/px.
Observations were acquired at four different pointings in order to cover the inner
region of the jet (see \oiii\ frame in Fig.~\ref{F-neb}). 
All data were reduced using the ESO GIRAFFE pipeline v2.9.2, which comprises debiasing, flat-field correction and wavelength calibration.
Individually reduced exposures were then combined to produce a master data cube for each pointing.
A log of the VLT+FLAMES observations is presented in Table~\ref{T-specobs}.

\begin{table}
\caption{Log of the VLT+FLAMES spectroscopic observations. 
The first column is the start of the observing night. The second column is the Julian Date at mid point of all exposures from the same pointing. 
The third column shows the total integration time. The last column refers to the telescope pointing. 
See text for more details. 
}
\label{T-specobs}
\centering
\begin{tabular}{llrrl}
\hline\hline
Date &JD & Total  exp.    & POS \\
          &         & time (sec)   &      \\       
\hline
 2012-10-03 & 2456203.683866   &   30; 100; 570   & 1  \\
  & 2456203.699699    &   1535  & 2\\
  & 2456203.769827   &  1228   & 3 \\
  & 2456203.786574   &   1228 & 4\\
\hline
 \end{tabular}
\end{table}

%%%%%%%%%%%%%%%%%%%%%%%%%%%%%%%%%%%%%%%%%%%
%%%%%%%%%%%%%%%%%%%%%%%%%%%%%%%%%%%%%%%%%

\section{Image processing}\label{S-pr}

To allow a careful study of the proper motions of nebular features over the whole monitoring period, 
it was necessary to map all images to the same reference frame.

The first step was to find the astrometry solution of each frame. Owing to
the lack of suitable field stars near R~Aqr (particularly problematic in short exposures taken with very narrowband filters), the
field geometric distortions and rotation in the NOT and VLT frames were
computed from more populated fields observed with the same filters.
Those fields were observed as close to the R Aqr observations in time
as possible in order to minimise the impact of any long-term effects.
The analysis allowed us to conclude that the adopted filters do not
introduce additional geometric distortions and that a common
astrometric solution can be obtained.
Astrometric solutions were calculated using the tasks {\it ccmap} and
{\it ccsetwcs} in IRAF and applied to the R Aqr data using the Python
based Kapteyn Package \citep{KapteynPackage}.
Errors in the astrometric solutions were on average $0''.19$ $\pm$ $0''.07$. No less
than 50 stars per field were used, occasionally up to 200.  For the NTT
1991 images it was not possible to obtain a precise astrometric solution at this stage.

As a second step, all images were matched against
a reference frame in all filters (data with $^\ast$ in
Table~\ref{T-imaobs}) using the stars in the FOV other than the R Aqr central
star.  This step was needed in order to place all frames onto the same
pixel scale for a direct pixel-to-pixel comparison. 
The IRAF tasks {\it geomap} and {\it geotran} were used to perform the
matching.  With this step, all the frames were resampled to the scale
of the reference frame, namely $0''.19$ pix$^{-1}$.
The RMS of the matching for \oiii, \ha+\nii, and \nii\ images was
mostly smaller or equal to $0''.07$. 
For the 1991 \ha+\nii\ image, the matching errors were
$\sigma_{\mathrm{RA}}=0''.12$, $\sigma_{\mathrm{DEC}}=0''.11$.
In \oii{}, only a few field stars are available, and therefore we adopted the 
matching solution from the \oiii\ frame obtained on the same night.  
A posteriori check using the few field stars available suggests a matching error
generally below $0''.08$ for this filter.
In the case of the 1997 \oii\ frame, there were no \oiii\ observations acquired, and 
consequently a solution was found directly from the five field 
stars available.  The matching errors were $\sigma_{\mathrm{RA}}=0''.09$,
$\sigma_{\mathrm{DEC}}=0''.12$.

Considering the large time spanned by our data, the possible influence
of proper motions of field stars used for astrometry was also investigated. Proper motions
were taken from the USNO-B1.0 catalogue \citep{2003AJ....125..984M}, or
from the UCAC4 catalogue \citep{2013AJ....145...44Z} if measurements from the former were not
available. We conclude that the corresponding errors in the matching
of the images are negligible.

A correction for the proper motion of the central
binary star of R~Aqr itself, which is non-negligible over the period
considered given the proximity
of the system, was then applied. We measured the proper motion directly in all non-saturated images, while the value of the 
proper motion from UCAC4 ($\mu_{\mathrm{RA}}=29.5$ mas yr$^{-1}$, $\mu_{\mathrm{DEC}}=-32.1$ mas yr$^{-1}$) was used for the saturated frames.  
Finally, all images were 
aligned with respect to the central star using the 
frame obtained  on September 5, 2007 as a reference.

%%%%%%%%%%%%%%%%%%%%%%%%%%%%%%%%%%%%%%%%%%%
%%%%%%%%%%%%%%%%%%%%%%%%%%%%%%%%%%%%%%%%%%%%%%%

\section{Overall properties of the R~Aqr outflows}\label{S-of}

% [width=4.5cm] will fit them all in one row.
% [width=4.5cm] in two rows 2 by 2.
\begin{figure*}
\centering
\includegraphics[width=8.5cm]{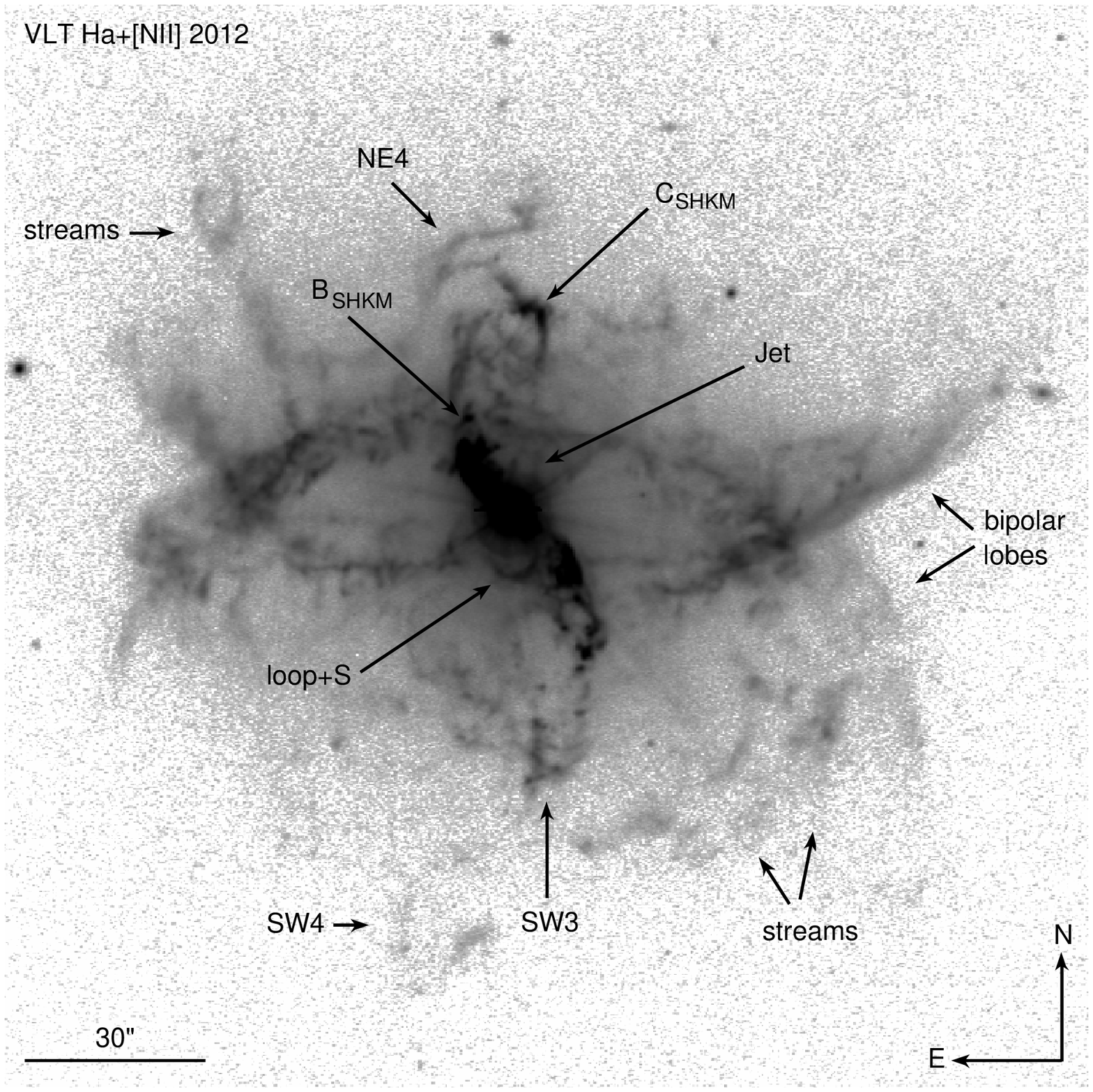}
\includegraphics[width=8.5cm]{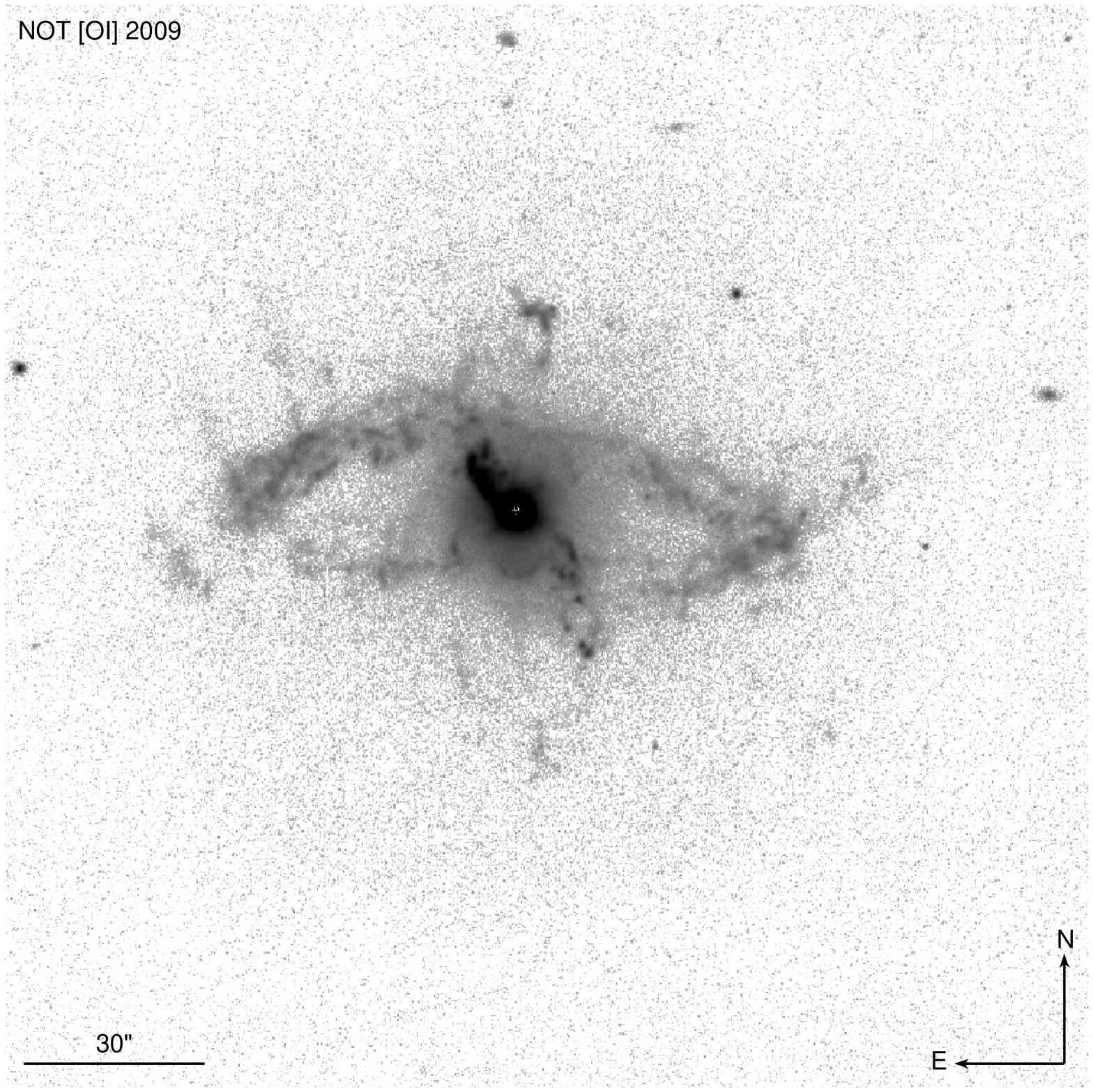}
\includegraphics[width=8.5cm]{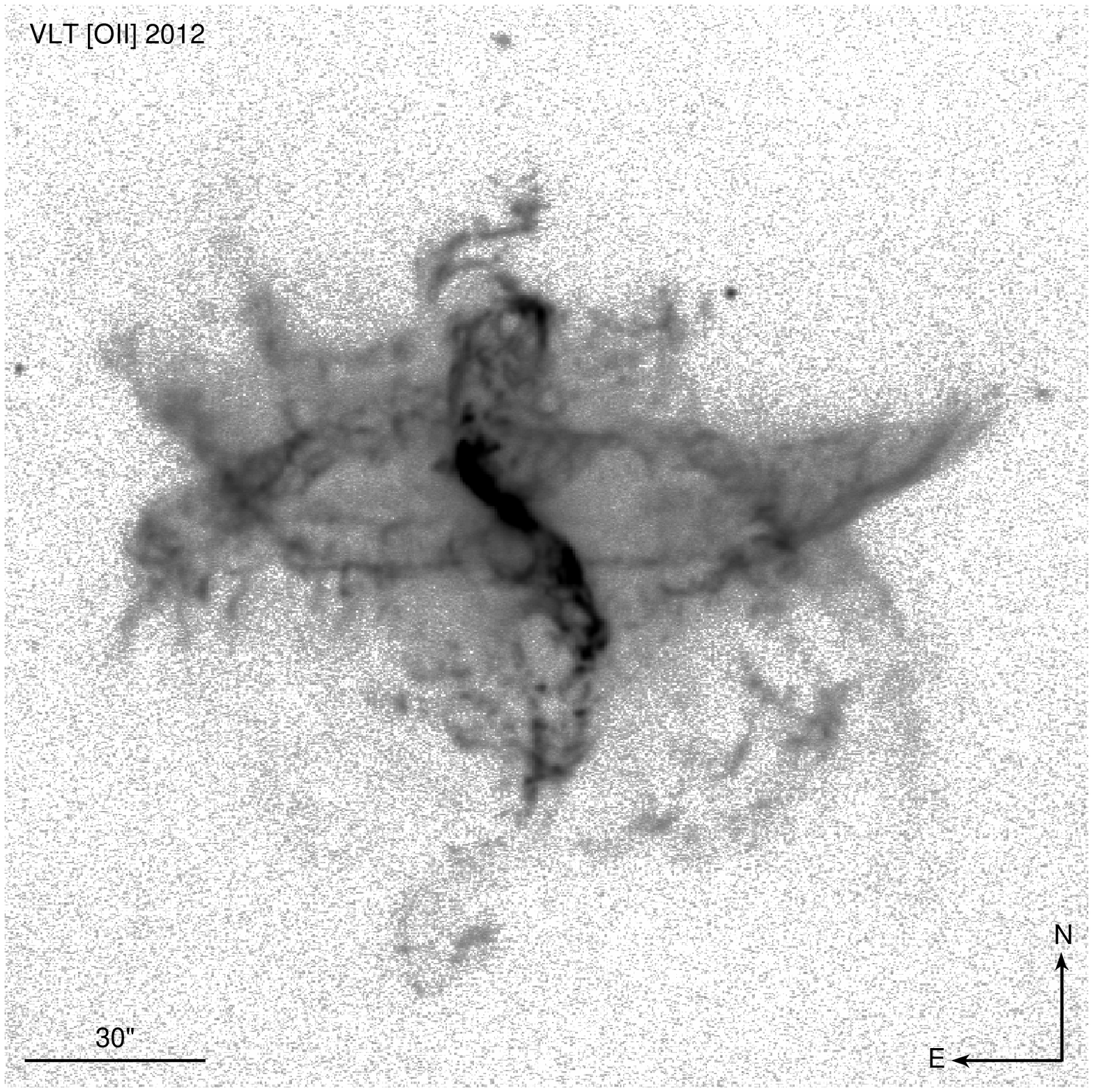}
\includegraphics[width=8.5cm]{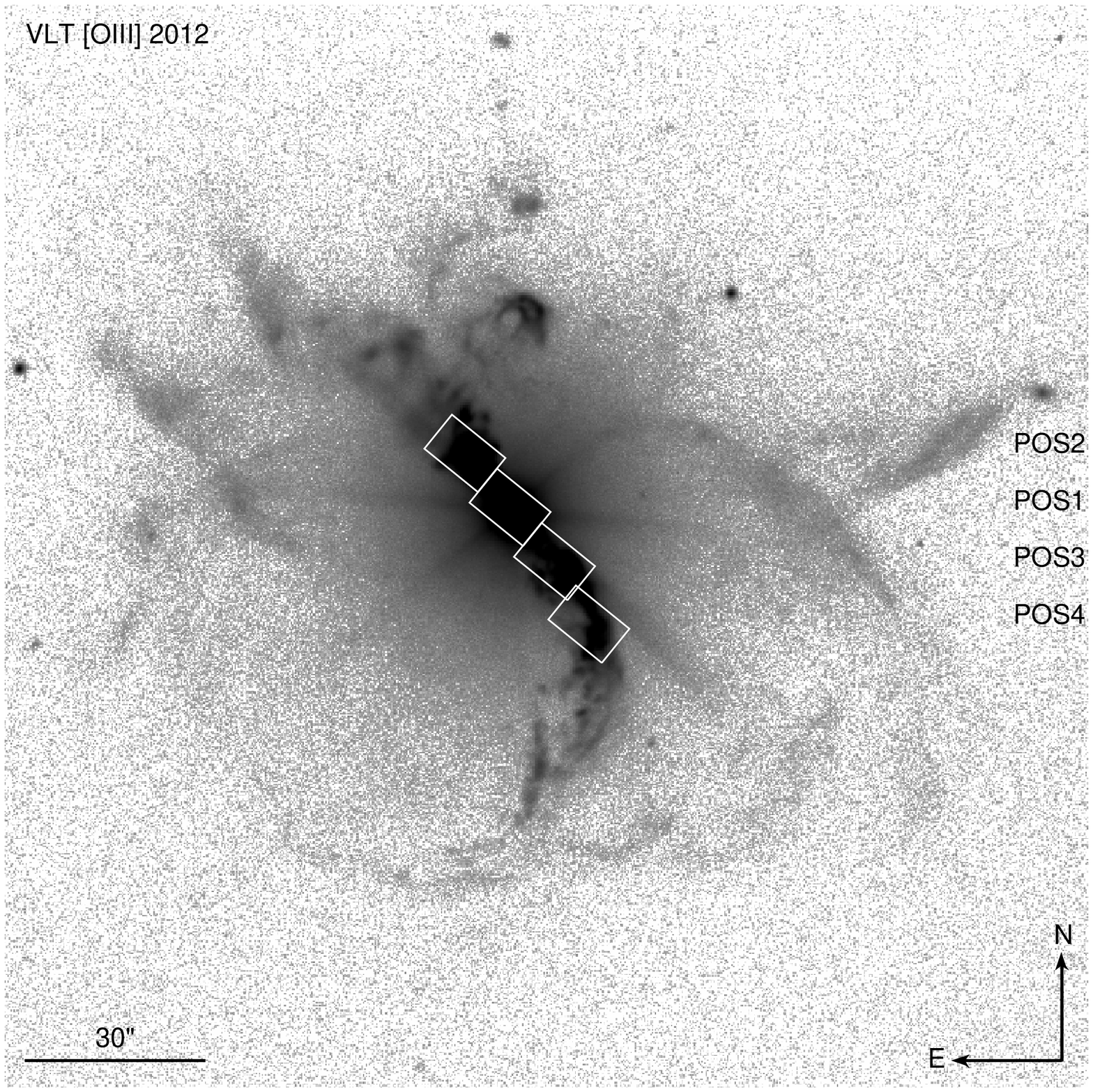}
\caption{VLT 2012 \ha+\nii, \oii, \oiii\ and NOT 2009 \oi\ frames. On the \oiii\ frame the white boxes represent the spectroscopic observations 
(see Section \ref{S-sp}). The FOV of all frames is $3'\times3'$. North up, East left.} 
\label{F-neb}
\end{figure*}
% see swo3.ps in Figures/SW_o3/old_larger

The R Aqr outflow consists of a bright, curved jet structure contained within a larger bipolar nebula \citep{1990ApJ...351L..17H}. 
The inner regions of the jet extend along the northeast-southwest direction, at an apparent angle of 40 degrees  from the 
symmetry axis of the bipolar lobes which instead are very close to the north-south orientation. 

\citet{1985A&A...148..274S} also identified an inner
bipolar nebula, oriented along the same symmetry axis as the larger
lobes but on a smaller scale. However, this latter structure is not obviously visible in any
of our images. Fig.~\ref{F-neb} shows the large scale morphological properties of
these outflows and the faint new details revealed by our deep imaging.

%%%%%%%%%%%%%%%%%%%%%%%%%%%%%%%%%%%%%%%%%%%%%%%

\subsection{Structure and expansion of the bipolar nebula}
\label{S-neb}

The large system of lobes and their bright equatorial waist are
mainly visible in the lower-ionisation light of \ha+\nii, \oii, and
\oi\ (Fig.~\ref{F-neb}). Indeed, \citet{2003ASPC..303..423G} show large
\nii/\ha\ and \sii/\ha\ flux ratios in the ring of the bipolar nebula
that they ascribed to shock ionization.

The appearance of the ring and lobes, at the high resolution provided
by our images, is complex. In the light of low-ionization ions, the ring is broken
into knotty and filamentary structures, while it looks much smoother in the
light of higher ionization species such as \oiii.   Furthermore, our deep images show for the first time fainter features 
that we designate as streams in Fig.~\ref{F-neb}. 
Streams appear in the NE and SW direction of the bipolar lobes extending up to $1'.2$ from
the central star. On the Southern side, the streams seem to replicate the curved
appearance of the outer regions of the jet.

The overall structure of the bipolar nebula does not show notable
changes over the 21 years considered (1991 - 2012). We therefore use the
so-called "magnification method" \citep[see][]{1999AJ....118.2430R,2007A&A...465..481S} 
to calculate the expansion in the plane of the sky and hence 
the age of this structure.  The essence of
the method is to find the magnification factor, $M$, that cancels out
residuals in the difference image obtained by subtracting the first
epoch, magnified image from the second epoch image. The method assumes
homologous expansion of the structure, but also allows one to identify
deviations from this assumption.

%\section{Expansion of the bipolar nebular}\label{S-age}

%macro read magnif.macro mneb, analyses_magnif.dat
The magnification factor was found using the 1991 and 2012
\ha+\nii\ images.  Owing to the different instruments, filters
properties, and observing conditions, we matched the point-spread function of images using field stars, 
and we also rescaled them in brightness using a portion of
the nebula itself.
Increasing magnification factors were then applied to the 1991 frame,
and the differences with the 2012 image computed. A precise
determination of the best-fitting magnification factor is limited by
the highly inhomogeneous morphology of the nebula. By dividing the nebula into quadrants, we 
found that the best-fitting $M$ values were as follows:
NE 1.030, SE 1.033, SW 1.033. 
Due to the faintness of the NW region of the nebula no measurement could be obtained 
from that quadrant, and for this reason we also 
defined a western (W) region restricted to the prominent part of the waist in that direction 
where a value of 1.033 to 1.036 was derived.

Assuming that the nebula has grown in time at constant velocity, the
resulting nebular age, computed with respect to the first epoch, is $T = \Delta t / (M-1)$, where $\Delta t$ is
time lapse between the two epochs, in our case 21.17 years.
%21.1690115
The $M$ values above imply an average age, weighted by errors,  of the bipolar nebula of
$T_\mathrm{{bip}}=$$653\pm35$ years in 1991, which compares well with previously 
published results (see Section \ref{S-intro}).  
\cite{2005A&A...435..207Y} found that R Aqr could have experienced a nova explosion in A.D. 1073 
and A.D. 1074. Without knowing the initial conditions of the explosion and of the circumstellar medium (ISM) 
it is not possible to find further support to the relation between the bipolar nebula and the possible 
ancient nova outburst, nor to discard it.

The estimated ages of the SW streams indicate that they are likely part of the extended nebula, 
rather than jet features.
The stream in the NE is not detected in our 1991 \ha+\nii\ image therefore no age estimate can be given. 

\subsubsection{New faint outer features}
\label{S-new}
 
\begin{figure}
\centering
\resizebox{\hsize}{!}{\includegraphics{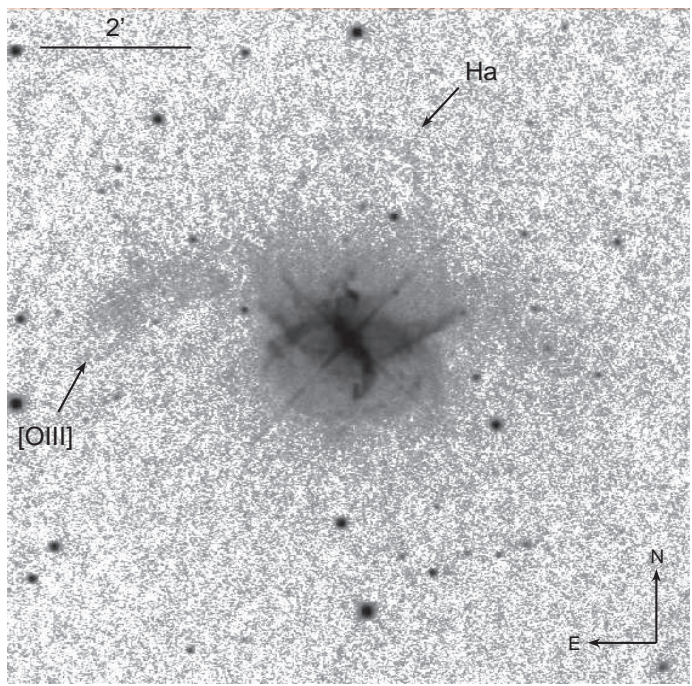}}
\caption{Terroux image showing faint outer \oiii\ and \ha\ features. FOV=$9'\times9'$. See text for more details.
\label{F-sid}}
\end{figure}

An additional set of images acquired in Terroux Observatory reveal new faint outer features in \oiii\ and in \ha. 
A combined \oiii\ and \ha\ image is presented in Fig.~\ref{F-sid}. It reveals the existence of a thick \oiii\ arc along the east-west direction 
with an extent of $6'.4$, and a thinner and fainter \ha\ loop, which extends to the North up to $2'.8$ from the central source. 
The latter may be related to the streams described in the previous section.

The \oiii\ arc is confirmed by stacking up all our other \oiii\ and \oii\ long exposure frames, but at a low signal-to-noise ratio.  
Unfortunately, the \ha\ feature is undetected in our VLT images, as their FOV does not fully cover the region.
It is likely that these features are related to the mass loss from the red giant and/or a nova eruption from the white dwarf 
in the earlier evolutionary stages of the system.

%%%%%%%%%%%%%%%%%%%%%%%%%%%%%%%%%%%%
%%%%%%%%%%%%%%%%%%%%%%%%%%%%%%%%%%%%

\subsection{Kinematic distance}
\label{S-d}

% see analyses_magnif.dat "ELLIPSE"
The combination of our determination of the apparent expansion of the
bipolar nebula with the radial velocity measurements of 
\citet{1985A&A...148..274S} allows us to derive the expansion parallax
of R~Aqr nebula. In particular, \citet{1985A&A...148..274S} found
that the equatorial waist can be modelled as an inclined ring
expanding at a speed of $V_\mathrm{{exp}}=55$~\kms.

The angular expansion of matter along the major axis of the projected
ring, over the period $\Delta t$ considered, is 
\mbox{$\Delta \alpha = \alpha (M-1)$}, where $\alpha$ is the distance from the central
source of the intersection of the ring with the plane of the sky.
Knowing the linear speed $V_\mathrm{{exp}}$, the distance to R~Aqr
follows immediately from the relation (in convenient units)

\begin{equation}
D (\mathrm{pc}) =  0.211 \frac{V_{\mathrm{exp}} (\kms) \Delta t (\mathrm{yr})}{\Delta \alpha ('')}
\end{equation}

Adopting the average value of the magnification factors determined
above, and fitting an ellipse to the nebular ring to measure its major
axis, we obtain a kinematic distance to R~Aqr of 178$\pm$18~pc. 
In general, this kinematic distance is in fair agreement with previous estimates. 
The nebular kinematics were first used by \citet{1944MWOAR..16....1A} to derive a distance of 260pc, later revised down 
to 180--185pc \citep{1985A&A...148..274S}.  This value is also in good agreement with the estimate of 181pc 
by \citet{1978ApJ...225..869L} based on the absolute magnitude of R Aqr at 4$\mu$m and an assumed value of  
$-$8.1$^{m}$.  Hipparcos parallax measurements by \cite{1997A&A...323L..49P} result in a slightly larger 
distance of 197pc, in strong agreement with the estimate based on the separation of the orbital components measured 
by the VLA \citep{1997ApJ...482L..85H}.  More recently, parallax measurements of SiO maser spots using VERA have 
indicated a yet greater distance of 214--218pc \citep{2010A&A...510A..69K,2014PASJ...66...38M}.

%%%%%%%%%%%%%%%%%%%%%%%%%%%%%%%%%%%%%%%%
%%%%%%%%%%%%%%%%%%%%%%%%%%%%%%%%%%%%%%%%

\section{Structure and expansion of the jet}
\label{S-j}

\begin{figure}
\centering
\resizebox{\hsize}{!}{\includegraphics{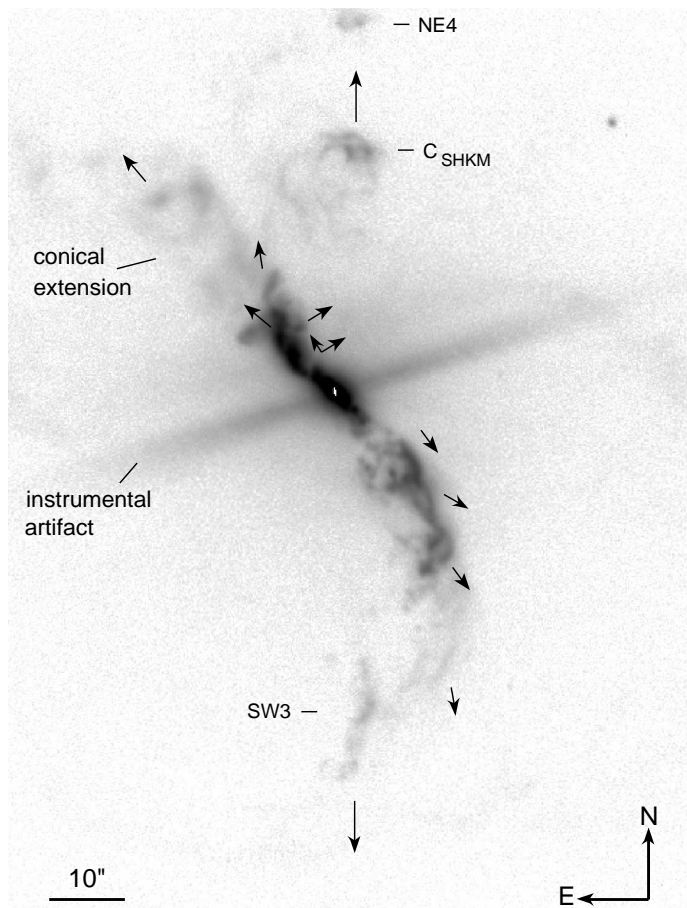}}
\caption{NOT 2007 \oiii\ image, with logarithmic display cuts that
  highlight the overall structure of the jet. Arrows point the expansion direction. FOV is $1'.5\times2'.0$.
\label{F-noto3}}
\end{figure}

Unlike the large-scale bipolar nebula, which within our present errors in the determination of the apparent 
and radial motions is well modelled assuming a mainly ballistic expansion, 
the jet shows a much more complex and irregular evolution. In our images, features identified by previous authors have brightened or faded, or 
even broken into multiple components moving along different directions. The bulk motions of various regions of the jet determined using 
our images are highlighted in Fig. \ref{F-noto3}. They demonstrate that, while the overall flow 
pattern is  consistent with a radial expansion of the jet, in some regions there are significant deviations, even perpendicular the radial direction of expansion. 
Their complex changes of appearance are  probably the combination  of illumination/ionisation variations as well as shocks. This makes the 
magnification method inappropriate to describe them. To illustrate this, we will discuss the evolution of individual 
features across our multi-epoch imagery in the following subsections.

Figs.~\ref{F-ne} and \ref{F-sw} show the evolution over time of the northern and southern region of the jet in the most relevant epochs and filters.
Note that the bright central area is often saturated, with strong charge overflow especially in \ha+\nii. 
The intensity level of the different frames in both figures was adapted to maximize the visible information.
The small blob near the central source in the \oii\ 2007 and 2009 frame,
pointed to with arrows in Fig.~\ref{F-ne}, are a red-leak images of
the central star (displaced because of atmospheric differential
refraction at the significant airmass of these observations). 
In Fig.~\ref{F-sw} on the 2009 \oiii\ frame, instrumental artifacts are visible as faint lighter thick lines 
emanating from the central star at PAs 135${\degr}$ and 225${\degr}$.

In order to avoid confusion, some clarification of the nomenclature of the jet features used in the past is in order. 
We refer to  the earliest designated jet features from \citet[][hereafter referred to as SHKM]{1982ApJ...258L..35S} as 
B$_{\text{SHKM}}$, C$_{\text{SHKM}}$, and the ``loop''.
Later publications revealed 
more details of the jet closer to the central source, which were designated as knots A and B by 
\citet{1983ApJ...267L.103K}, and knot D and S by \citet{1988ApJ...329..318P}. Knot S is 
a brightness concentration inside the ``loop'' feature and is therefore often referred to as loop+S. 
We keep the designations of the later authors. The aforementioned features, as well as the new ones found in this work, 
are marked in Figs.~\ref{F-neb}, \ref{F-noto3}, \ref{F-ne}, and \ref{F-sw}. 

Most of our images, as well as previously published data (imaging and spectra), show that the
NE jet is brighter than the SW jet at both large scale and in the central area 
\citep{1994A&A...287..154P}. However, our short exposure \oiii\ and \oii\ VLT images from
2012 reveal that in the central region \mbox{(< $2''$)}, the SW jet is brighter. 
This is also confirmed by our spectral data from 2012 (Fig.~\ref{F-rv3}). This seems to have appeared 
a few years before 2012, as the images from 2002 and 2007 show equal brightnesses for
NE and SW in the central area, while by 2009 and 2012 clearly the SW is more prominent.
This central area is resolved in the recent high spatial resolution SPHERE images from
2014 by \cite{2017A&A...602A..53S}. They also detect that the SW jet is brighter than its
NE counterjet in these central areas. 

%%%%%%%%%%%%%%%%%%%%%%%%%%%
%%%%%%%%%%%%%%%%%%%%%%%%%%%

\subsection{The NE Jet}
\label{S-nej}

\begin{figure*}[!th]
\centering
\includegraphics[width=16cm]{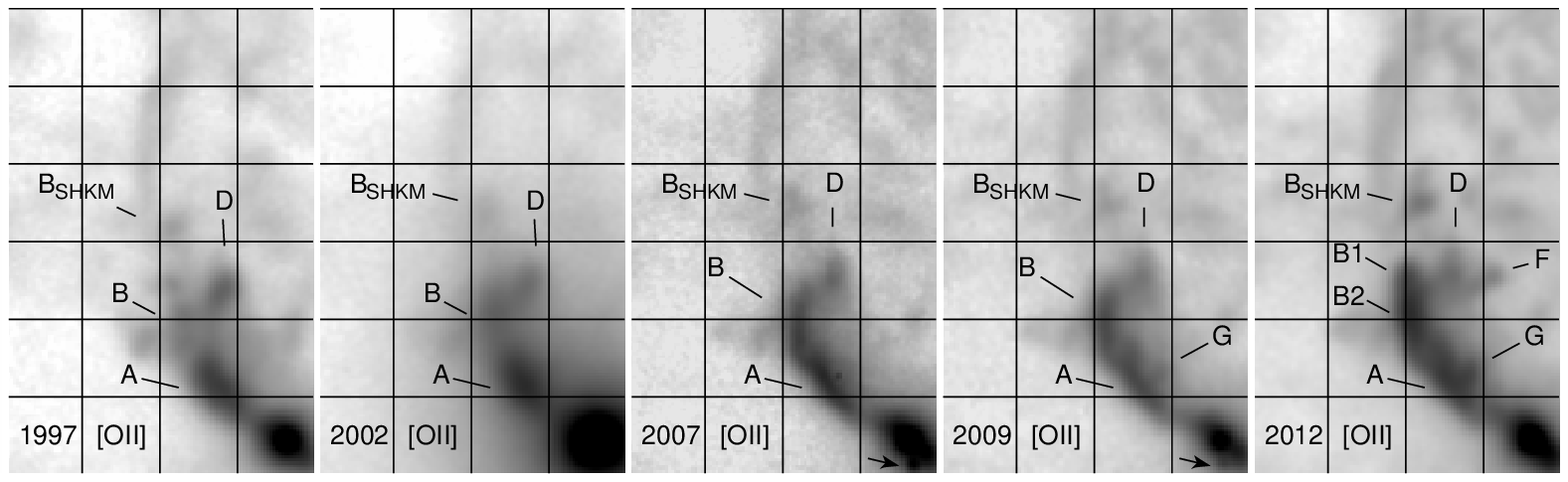}
\includegraphics[width=16cm]{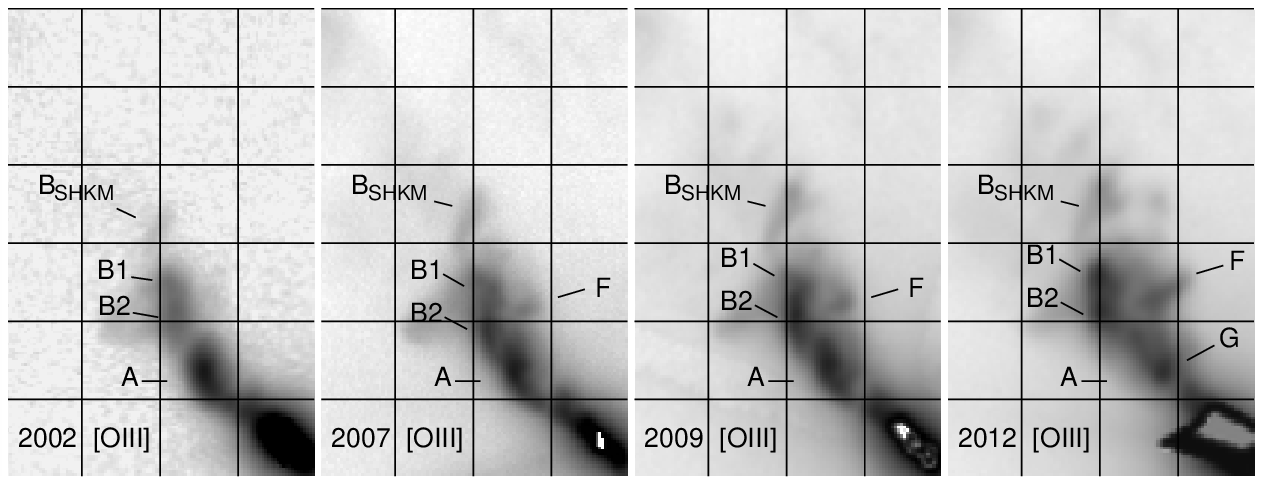}
\includegraphics[width=16cm]{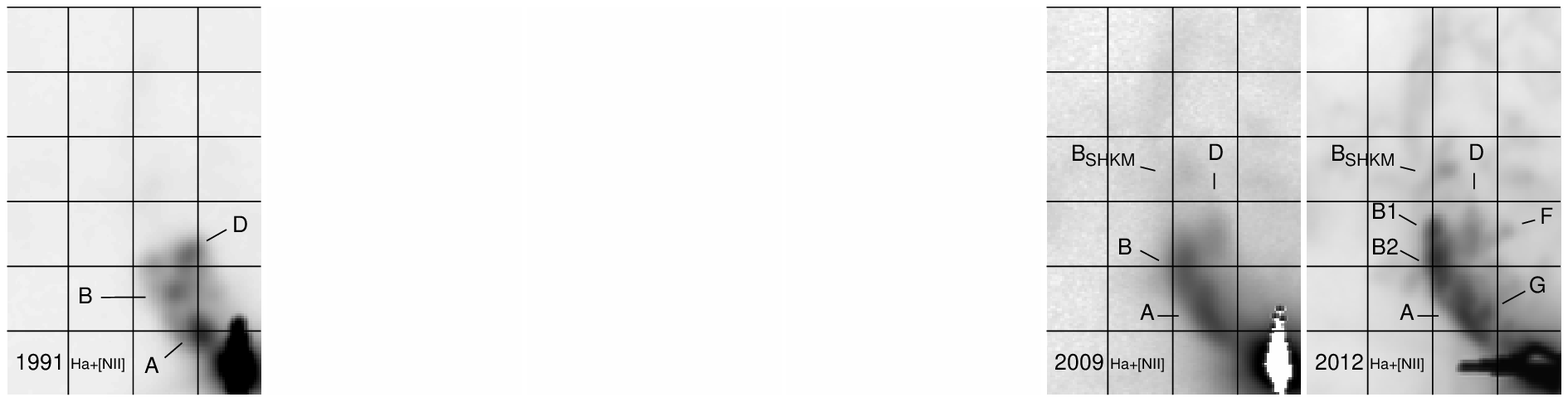}
\caption{NE jet in \oii, \oiii, and \ha+\nii\ frames from top to bottom respectively. 
One square is $5''\times5''$. The FOV of each frame is $20''\times30''$. North is up, East to the left. \label{F-ne}}
\end{figure*}

The evolution of the northern, bright part of the jet is shown in Fig.~\ref{F-ne}. It has a complex, knotty, and variable appearance. 
For instance, the feature that we name as F first appeared in 2007 in the \oiii\ filter. 
The 2012 image indicates that, contrary to the general radial expansion, it is seemingly moving towards the west. 
The same applies to feature G, which appeared in 2011. Their westward lateral movements  
are indicated by the two arrows in Fig. \ref{F-noto3}. Transformed into linear velocities, their motions would imply speeds of between 500 and  900 \kms. 
This is several times larger than the bulk radial motions from imaging and spectroscopy, indicating that very likely they do not reflect the 
true physical movement of a clump of material but rather are due to changes in the ionisation conditions of the region.

At a distance of $\sim20''$  (see Fig.~\ref{F-noto3}) the northern jet splits into two components, a brighter one bending toward the north
and ending in filamentary features such as C$_{\text{SHKM}}$ and NE4, and a
fainter, more diffuse one that seems a straight, conical extension of the innermost jet.
The latter is only visible in the \oiii\ emission line and its cone-like morphology suggest that it may be an illumination 
effect \citep{2011A&A...529A..43C}, further indicating that the role of changing illumination/excitation is critical in understanding the structure of the jet. 

\citet{1991ApJ...370..590B} used optical emission line ratios to conclude that the R Aqr 
jet features are best fitted with local shock-wave models, as was firstly suggested by \citet{1985A&A...148..274S}. 
\citet{1991ApJ...370..590B} found that the brightening of knot D (and fading of knot B) came some 15 yrs after the 
brightening of knot B (in the late 1970s), concluding that the shock wave was propagating outwards at 90--100 \kms{}.  
As such, they predicted that the knot D should fade analogously 15 yrs after it was observed to brighten, which would approximately occur in 2003.
Inspection of our \ha+\nii\ data, Fig.~\ref{F-ne}, reveals that in 1991 B and D have similar brightness 
(just as in observation from 1986 in \citet{1991ApJ...370..590B}). Our next \ha+\nii\ frame is from 2009 were 
knot D, indeed, is much fainter than knot B, while at the same time the knot A is the brightest. By 2012, 
the relative brightness of knot B is around that of knot A, while knot D is the faintest. A similar tendency is detected in our 
\oii\ data where by 1997-2002 knot B has almost disappeared, while knot A and D are clearly brighter. 
By 2007 knot B starts to brighten again and quite soon (2009) becomes as bright as knot A. At the same 
time it does appear that knot D is fading. The similar relative brightening of knot B is visible in the \oiii\ frames. 
To conclude, the prediction of knot D fading is confirmed (though perhaps at a faster rate than predicted). 
In addition, the re-brightening of knot B may imply that another shock-wave is passing through the system. If so, it should eventually start 
again influencing knot D. However, from Fig.~\ref{F-ne} it is evident that, by our latest epochs, knot B 
has stretched as far out as knot D, but the latter is not brightening together with knot B. 

Over the years, features in the central area of the NE jet (Fig.~\ref{F-ne}) tend to get elongated along the expansion direction 
and eventually break into separate components. For instance, feature B breaks into B1 and B2. 
Similar stretching is happening with features A and D, which was seen to happen at 8 GHz radio band as early as 
1992 \citep{2004A&A...424..157M}. Our data shows that 
by 2012 in all filters feature A has a very elongated structure, which is clearly indicative of imminent breakup. 
The delay between optical and radio can be just due to the lower resolution available in optical wavebands.

As indicated with the arrows in Fig.~\ref{F-noto3}, the outermost NE jet features expand radially during the period considered. 
Therefore we use the magnification method (features  C$_{\text{SHKM}}$ and NE4, see Fig.~\ref{F-neb}) or measure 
directly the proper motion (feature B$_{\text{SHKM}}$, Figs.~\ref{F-neb} and \ref{F-ne}) to estimate their kinematical ages. 
The ages together with approximate distances from the central star are presented in the Table \ref{T-ad}. 
Ages are presented for the epochs 1991-07-06 and 2012-09-05. The distances are measured at the 2012-09-05 date 
because not all features were present on the 1991 image.  
Due to the extended nature of most of these features, distances from the central source are roughly estimated. 
Details related to the measuring can be found in Appendix \ref{A-jet}. Here only the main results are presented.

%see more from ima_analyse.dat
The measurements of the brightness peak of feature B$_{\text{SHKM}}$ indicate an age between 125 and 180 years. 
The average proper motion,  $\mu=0''.10\pm0''.02$ yr$^{-1}$, is compatible with the 
calculations in SHKM  ($0''.082\pm0''.014$ yr$^{-1}$), indicating a roughly constant expansion velocity over the last 50 years or so. 
Feature C$_{\text{SHKM}}$ was twice as bright in 2012 than in 1991. 
In SHKM, a slow brightness change for that feature is mentioned, but it is not clear if it was observed to be brightening or dimming. 
The age found for that feature, 286 yrs, is much younger than the bipolar nebula, though much older than the previously 
derived jet age of about 100 yrs \citep{1992Natur.355..705L,1993ApJ...411..235H}.
If the C$_{\text{SHKM}}$ feature is part of the jet, it is not surprising that the brightness has changed, as brightness variations 
have been seen among all features of the jet. 
The feature also demonstrates the jet's structural differences at different wavelengths, given that
C$_{\text{SHKM}}$ keeps its structure in time when 
considering single filter data, but its form varies in different filters. In \ha+\nii\ it has an arched shape, 
while in \oiii\ it seems a double arched or circular. In \oi\ it has a T-shape structure. 
For feature NE4, an age of $285\pm$61 years is found. 
This age is much younger than the bipolar nebula, though older than the jet, just as found for C$_{\text{SHKM}}$. 
Collectively our observations of B$_{\text{SHKM}}$, C$_{\text{SHKM}}$, and NE4 imply that they are real physical structures, 
and that their apparent morphological changes are not dominated by the changing ionisation.

\begin{table*}
\caption{Kinematic ages at epochs 1991-07-06 and 2012-09-05 together with an
 approximate distances for the epoch 2012-09-05 from the central source for the ballistic features of the jet.}
\label{T-ad}
\centering
\begin{tabular}{lllrcll}
\hline\hline
             & Feature      &  Age 1991 & Age 2012 & Distance & Method & Comments \\
             &                      &  yrs            &  yrs              &$''$            &  & \\       
\hline
North & B$_{\text{SHKM}}$& 125-180 &145 -- 200 &17  & direct  & Stable expansion velocity over the last 50 yrs,\\    % Filter for age: Ha+nii; oii; oiii  
                  &              &                        &                       &       &               & $\mu=0''.10\pm0''.02$ yr$^{-1}$\\ 
          &C$_{\text{SHKM}}$ & $286\pm12$  &$307\pm12$ &35 & magnif. &  Brightness variations. \\ 
                     &              &                        &                       &       &               & Structural changes in different wavelengths.\\ %age: Ha+nii  
          &NE4                          &$285\pm61$&$306\pm61$& 45& magnif.  & \\  % age:oii
\\  
South & loop+S&$160\pm40$&$183\pm40$&10 & magnif. & ``loop'' expanding steadily at least since 1960s.\\   % age:oii
           &              &                        &                       &       &               &   Significant brightness change of knot S.\\ 
         &SW3     &$215\pm36$&$236\pm36$&45 & magnif. & \\ %age:oii
         &SW4     &$880\pm150$&$900\pm150$&75 & magnif. & \\ % age:ha+nii
\hline
 \end{tabular}
\end{table*} 
% all the distances are taken from the OII

%%%%%%%%%%%%%%%%%%%%%%%%%%%
%%%%%%%%%%%%%%%%%%%%%%%%%%%

\subsection{The SW Jet}
\label{S-swj}

\begin{figure*}[!th]
\centering
\sidecaption
\includegraphics[width=16.0cm]{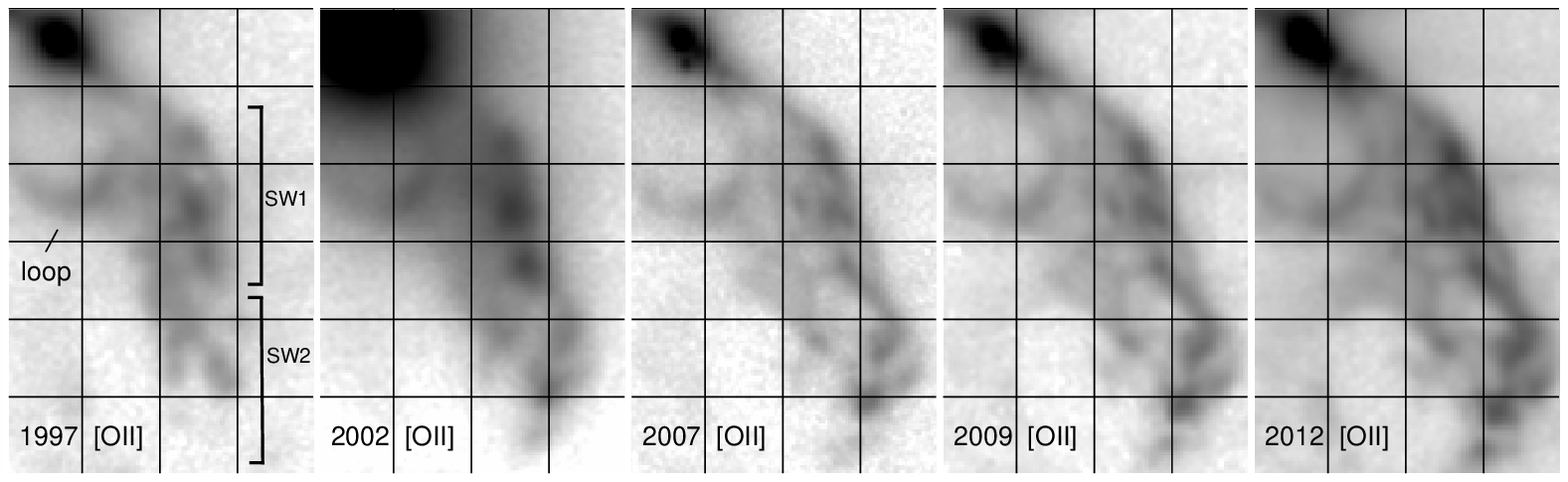}  % there is also 2011 but it is left out as it does not reveal any new features.
\includegraphics[width=16.0cm]{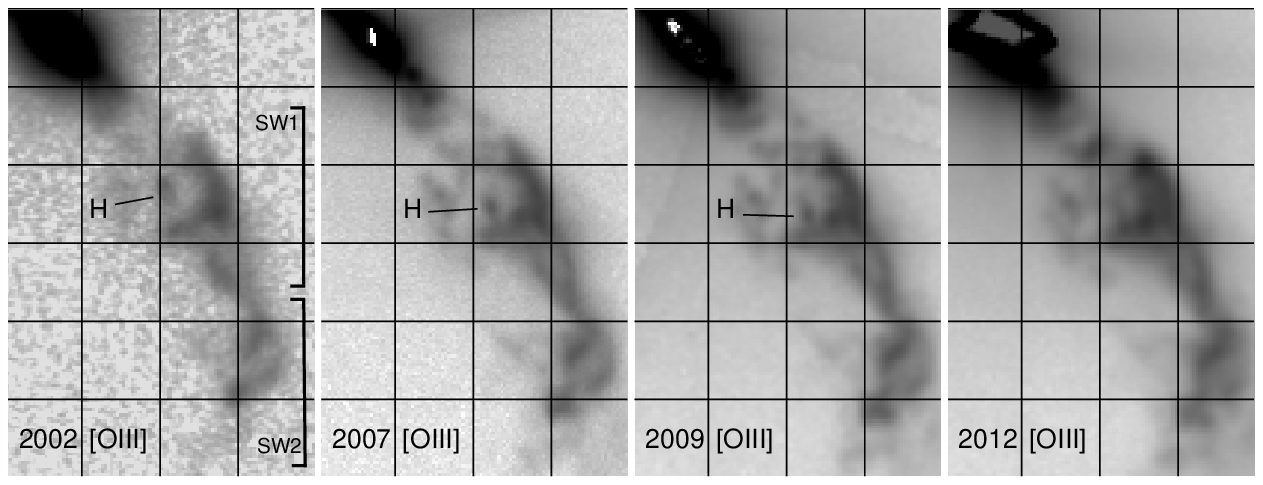}
\includegraphics[width=16.0cm]{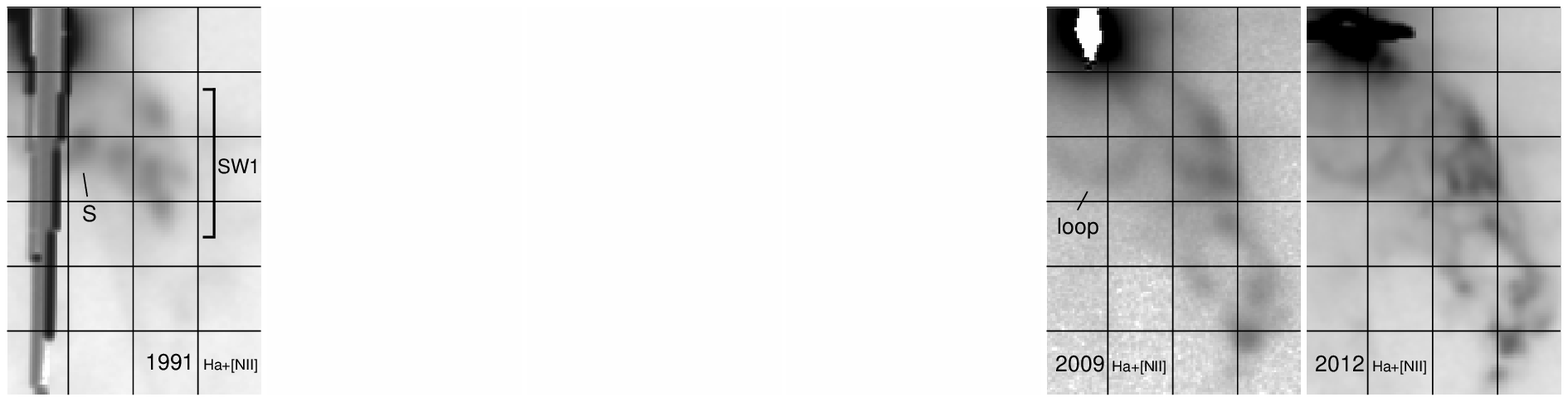}
\caption{SW jet in \oii, \oiii, and \ha+\nii\ frames. One square is $5''\times5''$. The FOV of each frame is $20''\times30''$. 
North up, East left. \label{F-sw}}
\end{figure*}

The evolution of the SW jet from 1991 to 2012 in the most relevant filters is presented in Fig.~\ref{F-sw}. 
We refer to the SW jet with the following nomenclature. SW1 consists of several blobs 
(apart from the loop+S) seen in the 1991 \ha+\nii\ frame in Fig.~\ref{F-sw}.
The rest of the SW jet visible in Fig.~\ref{F-sw} is named SW2. The more extended features SW3 and SW4 are 
also highlighted in Fig.~\ref{F-neb}. The expansion of the SW jet is more uniform than the northern one, showing 
mostly radial motions (features loop+S, SW4, and SW3 in Fig.~\ref{F-neb} 
and \ref{F-sw}).
However, based on our observations, a few interesting remarks can be made. 

First of all, the relative brightness of SW2, compared to SW1, has increased over the years. The complex was marginally visible in 
our 1991 \ha+\nii\ frame, and overall is now as bright as SW1. 
Significant structural changes are also detectable in the SW2 (compare 1997 and 2002 \oii\ in Fig.~\ref{F-sw}), and more structure has 
become visible which seems to connect SW2 with SW3 via faint curved filaments (see Fig.~\ref{F-neb}). 

% Feature H (previous K)
Feature H, shown in the light of \oiii\ in Fig.~\ref{F-sw}, is found to be moving faster than the surrounding jet 
at the epochs 2002, 2007, 2009, and then disappears or dissolves into the surrounding jet emission. 
% macro read intres.macro feature. Distance 178+-18
Its proper motion is on average $0''.30 \pm 0''.01$ yr$^{-1}$.
% The errors for the above averages are the STDEV of the two measurements.
Taking into account the distance to R Aqr found in Section~\ref{S-d} the average linear velocity would be $\sim$250 \kms. % 253.9575767 +- 7.315335233 kms
This is not as fast as the lateral movements detected in the NE jet but it is still faster than the jet in general, again possibly resulting from a 
shock-wave moving through the system rather than a real matter movement. 
If we assume constant velocity, it was ejected 38$\pm$1 years ago. 

In the SW jet, kinematical ages could be calculated using the magnification method for the radially expanding features loop+S, SW3, and SW4 
(see Figs.~\ref{F-neb} and \ref{F-sw}). 
As for the NE jet, details of the analyses are presented in Appendix \ref{A-jet} and Table \ref{T-ad}.
% loop+S
The feature ``loop'' preserves its horseshoe shape over our observing period in all filters in which it is detected. 
The brightness enhancement, knot S, was detected in 1986 by \citet{1988ApJ...329..318P} is still clearly visible in our 1991 \ha+\nii\ frame (see Fig.~\ref{F-sw}). 
At later epochs, the knot S becomes elongated along the loop until it almost disappears, but that part of the loop  
stays brighter than the other regions. Fig.~\ref{F-sw} also shows that the loop is slowly expanding towards the south.
The age found for the loop+S, 160 yrs, 
%giving an age about $170\pm40$ yrs at 1997.    
% 168.2267953  +-  37.38373228 years 
% see analyses_magnif.dat in /Users/TIINA/ASTRONOOMIA/RAQR/RAQR_1article/RAQR_analyse_magnif/Ha_NII_magnif
corresponds to an ejection event around the year $1827\pm40$. 
This is comparable with the ejection date $1792\pm32$ computed by SHKM. This also shows that no major changes have occurred 
in the evolution of the loop since their observations in the 1960s.
% SW3
The estimated age of feature SW3, $215\pm36$ years is much younger than the bipolar nebula, though older than the jet, just as found for NE4 and C$_{\text{SHKM}}$. 
% SW4
In the case of the newly-identified hook-shaped feature SW4 we find a rather large age ($880\pm150$ years) but as the feature is 
very faint, we conclude that its age is consistent within the uncertainties with that of the extended bipolar nebula.
% see analyses_magnif.dat in /Users/TIINA/ASTRONOOMIA/RAQR/RAQR_1article/RAQR_analyse_magnif/Ha_NII_magnif
The age of the ballistic features of the jet increases with the distance from the centre, consistent with continuous/repeated ejection events 
or that the more extended features have been slowed down by circumstellar material. 

%%%%%%%%%%%%%%%%%%%%%%%%%%%%%%%%%%%%%%%%%%%
%%%%%%%%%%%%%%%%%%%%%%%%%%%%%%%%%%%%

\section{Radial velocity measurements}\label{S-rv}

The \oiii\ 5007~\AA\ emission extends over the entire FOV of the ARGUS IFU pointings, except for very few spaxels. 
This resulted in $\sim$1200 usable individual spectra in total.  Radial velocity measurements were obtained from each lenslet by 
Gaussian fitting of the \oiii{} line using the {\it splot} task in IRAF. 
Every fit was visually checked and multiple Gaussians were used when needed.  The measured radial velocities were corrected first
to the Local Standard of rest (LSR), and then to R~Aqr systemic
velocity of -24.9 km~s$^{-1}$ \citep{2009A&A...495..931G}. 

\begin{figure*}
\centering
\includegraphics[width=4.5cm]{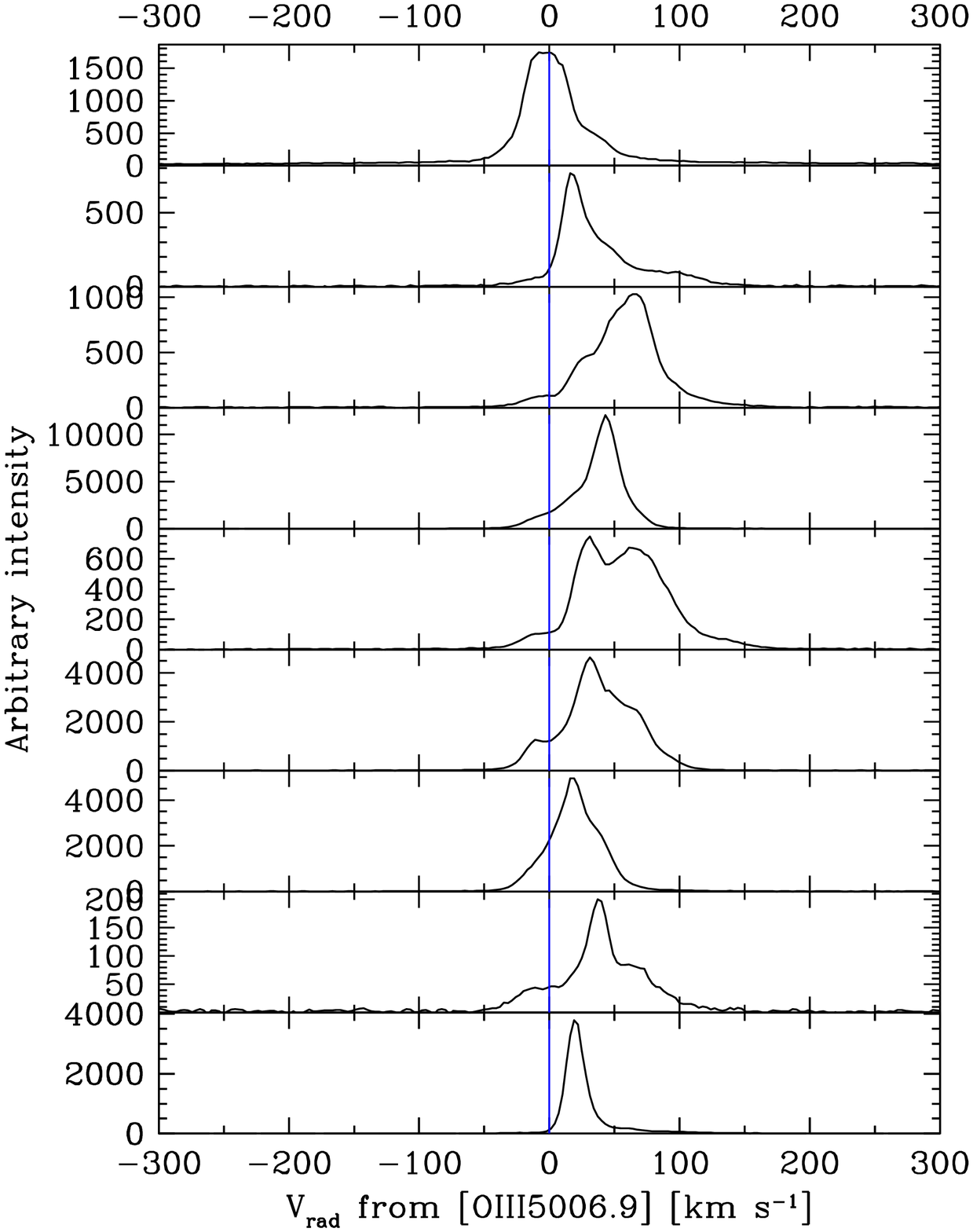}
\includegraphics[width=4.5cm]{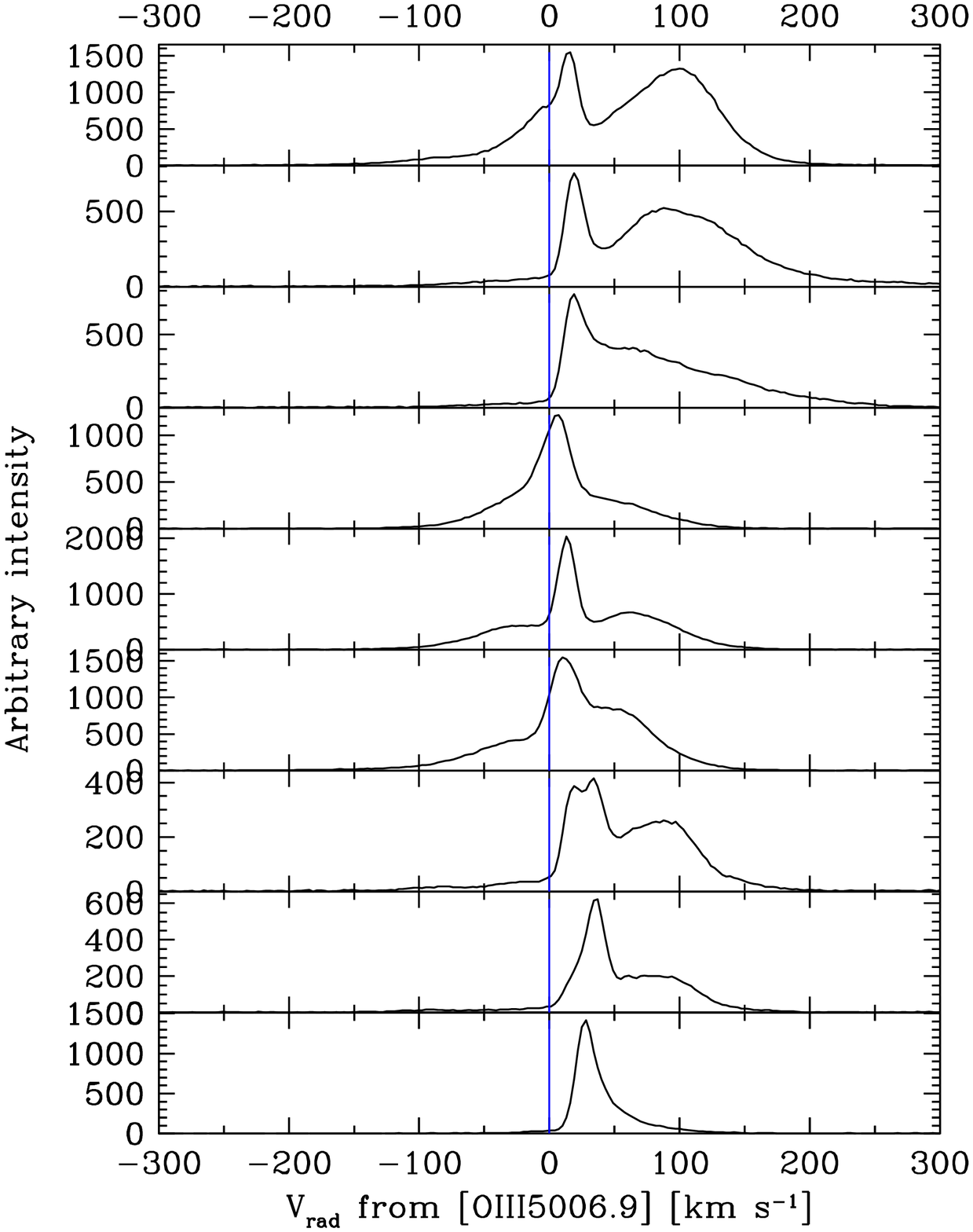}
\includegraphics[width=4.5cm]{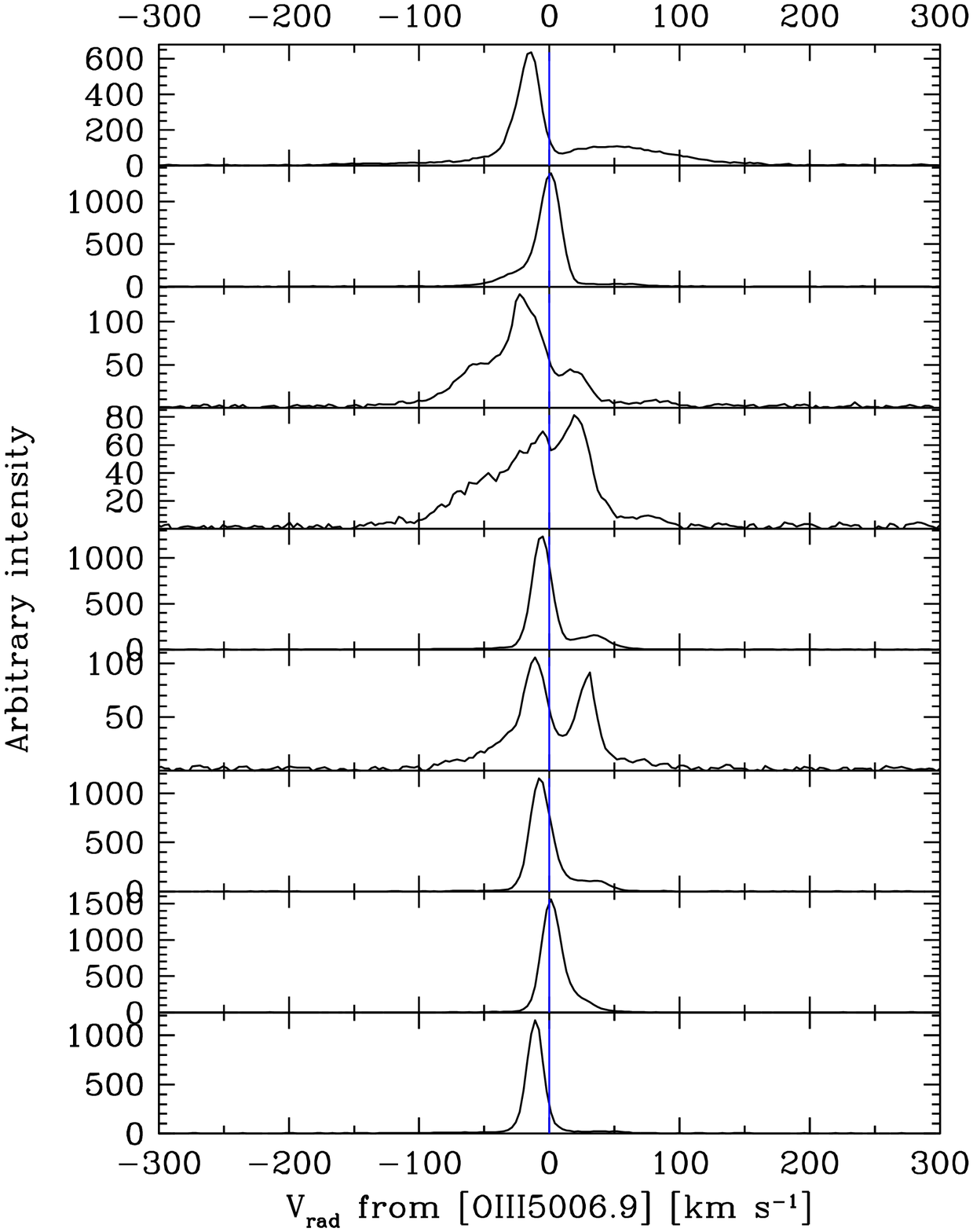}
\includegraphics[width=4.5cm]{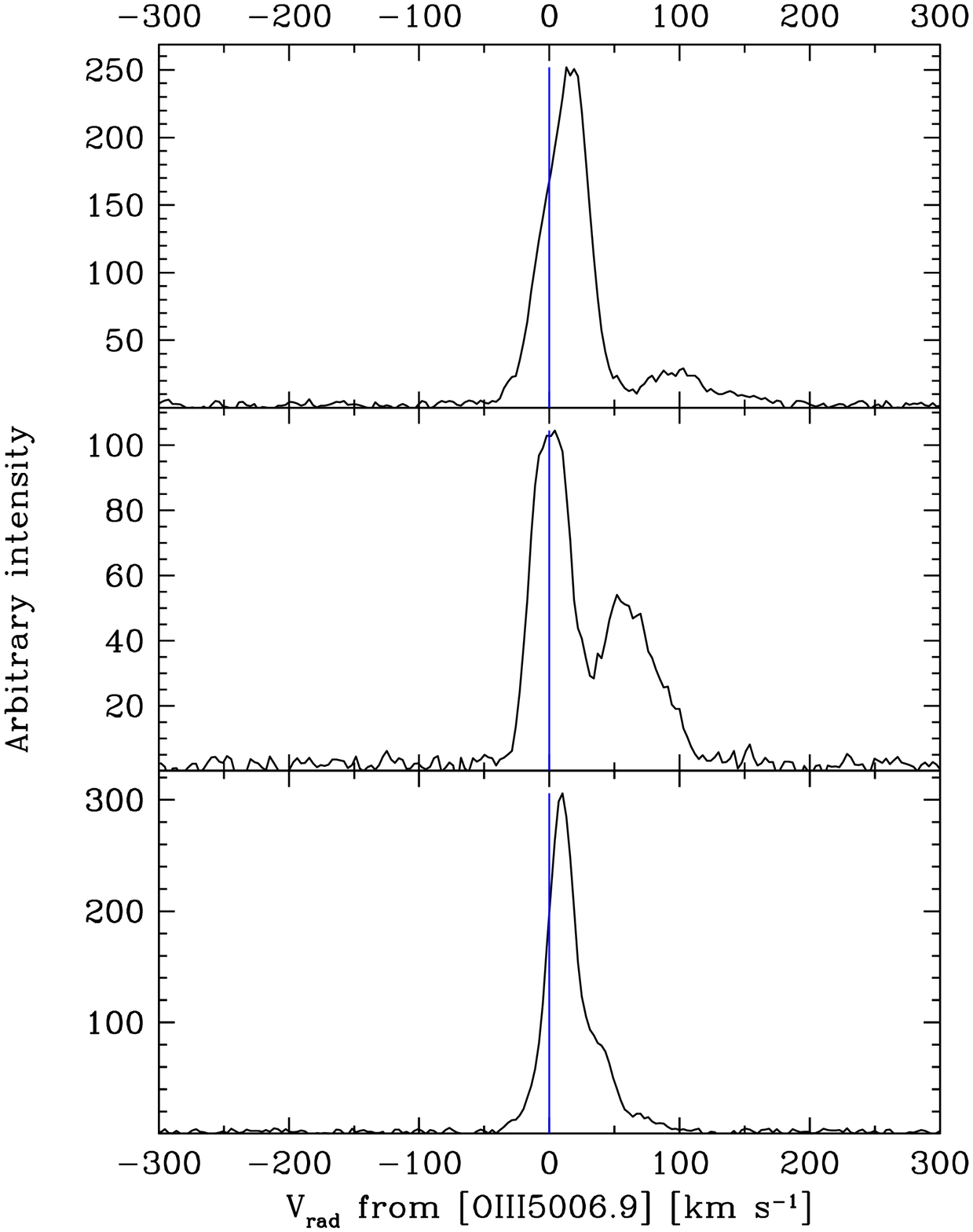}
\caption{Line profile examples from left to right POS1, POS2, POS3, and POS4. Blue vertical line 
refers to a radial velocity 0 \kms. All radial velocities are corrected for the systemic velocity.\label{F-lp}}
\end{figure*}

\begin{figure}
\resizebox{\hsize}{!}{\includegraphics{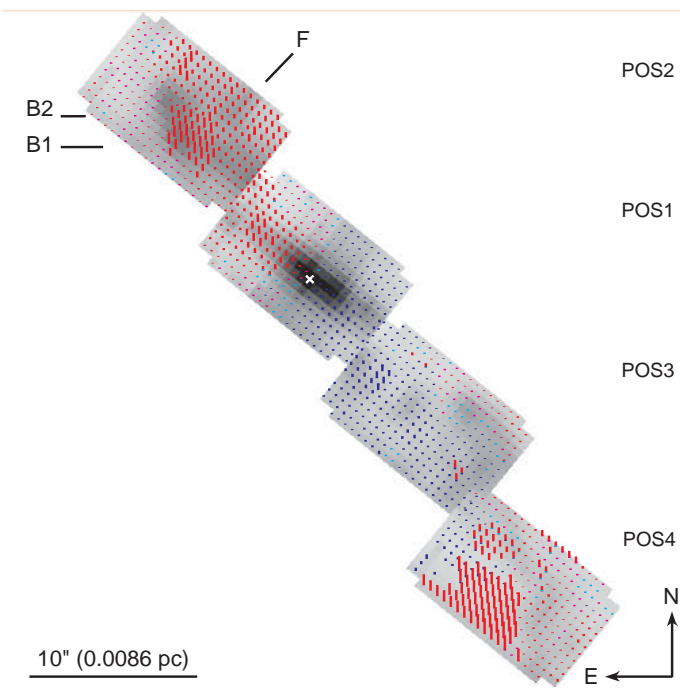}}
\caption{Radial velocities of the emission peaks, superimposed onto a 
greyscale representation of the reconstructed \oiii\ image from the 
IFU data. The FOV=$40''\times40''$. Radial velocities are indicated by lines, whose colour 
indicates if emission is blue- or red-shifted, and whose length is proportional to the absolute value of the velocity. 
Cyan and pink lines represent blue shifted velocities smaller than -5 \kms and redshifted smaller than +5 \kms respectively. 
The length of the cyan and pink lines are constant because otherwise they would be too small to be visible. 
The white X-point marks the position of the central star. The distance used for the linear size is taken from this work, 178 pc.
\label{F-rv3}}
\end{figure}

The \oiii\ line profiles are generally complex and sometimes display broad
wings, as illustrated in Fig.~\ref{F-lp}. We initially 
limit the discussion to the strongest emission peaks, defined as those spectral components whose integrated flux is 
larger than 75\%\ of any other component 
from the multi-Gaussian fit at each spaxel. Results are presented in Fig.~\ref{F-rv3}, where the radial
%see more from fits_help_aqr.dat argus.ps (macro read rv.macro allsky is part of it)
velocities of the strongest peaks are plotted on top of the reconstructed \oiii\ image.

At position 1 (POS1, see Fig.~\ref{F-lp}), that is 
centred on the star, the radial velocity varies from -33 to +72  % macro read rv.macro allsky using the file results_501_clean_flux.dat for all the position velocities.
\kms, with the innermost portion of the south jet being blueshifted,
and the north counterpart mainly redshifted. This trend continues 
further away from the centre, at position 2 (POS2, to the North) that is mostly redshifted (from -5 to +97 \kms) 
and position 3 (POS3, to the South), which is mostly blue-shifted with a few redshifted components (from -56 to +59 \kms). 
At position 4 (POS4), where the southern jet significantly bends, most of the emission becomes
redshifted, ranging from -38 to + 137 \kms. This general description is obviously complicated by the
very complex line profiles, which include additional emission
components as well as extended wings spanning a range of radial velocities as
large as $\sim$400 \kms.  % FWHM macro read rv.macro wstat   using the file results_501.dat
In general, however, it is clear that on the same side of the central star, both red- and blue-shifted regions are found, which - assuming 
purely radial motions - is a clear sign of a changing direction of the ejection vector crossing the plane of the sky. 
This in turn is usually associated to precession of the ejection nozzle as seen at large inclinations.

From Fig.~\ref{F-rv3} it is evident that broken up components of feature B, namely B1 and B2 (see Section~\ref{S-nej}), 
have different radial velocities, 
% featureHB1B2.numbers
$V_\mathrm{{rad_{B1}}}=+73\pm27 \kms$ and 
$V_\mathrm{{rad_{B2}}}=+22\pm22 \kms$, respectively. Similarly, we calculate their average FWHM to be $78\pm23 \kms$ 
and $32\pm24 \kms$, respectively. As such, B1 presents almost 3 times the radial velocity of B2 and nearly twice its velocity dispersion. 
From their motion in the plane 
of the sky, we see that the entire feature B (including components B1 and B2) has been moving steadily towards the NE (Fig.~\ref{F-noto3}), 
implying that their velocities in the plane of the sky should be similar.  As such, we can conclude that the true spatial velocity of B2 is significantly lower than that of B1.

Another feature for which we can determine a radial velocity from the data shown in Fig.~\ref{F-rv3} is feature F (see Section~\ref{S-nej}). 
% I consider 3x6 spaxels
It has a uniform radial velocity over the whole elongated feature, \mbox{$+36\pm2 \kms$}, with a narrow single gaussian 
line profiles FWHM $=16\pm1 \kms$. This would seem to imply that the feature F is moving almost completely perpendicular to the the 
line of sight, as its tangential velocity, 500 \kms (see Section~\ref{S-nej}), is much larger than the measured radial velocity.

\begin{figure*}
\centering
\includegraphics[width=12.55cm]{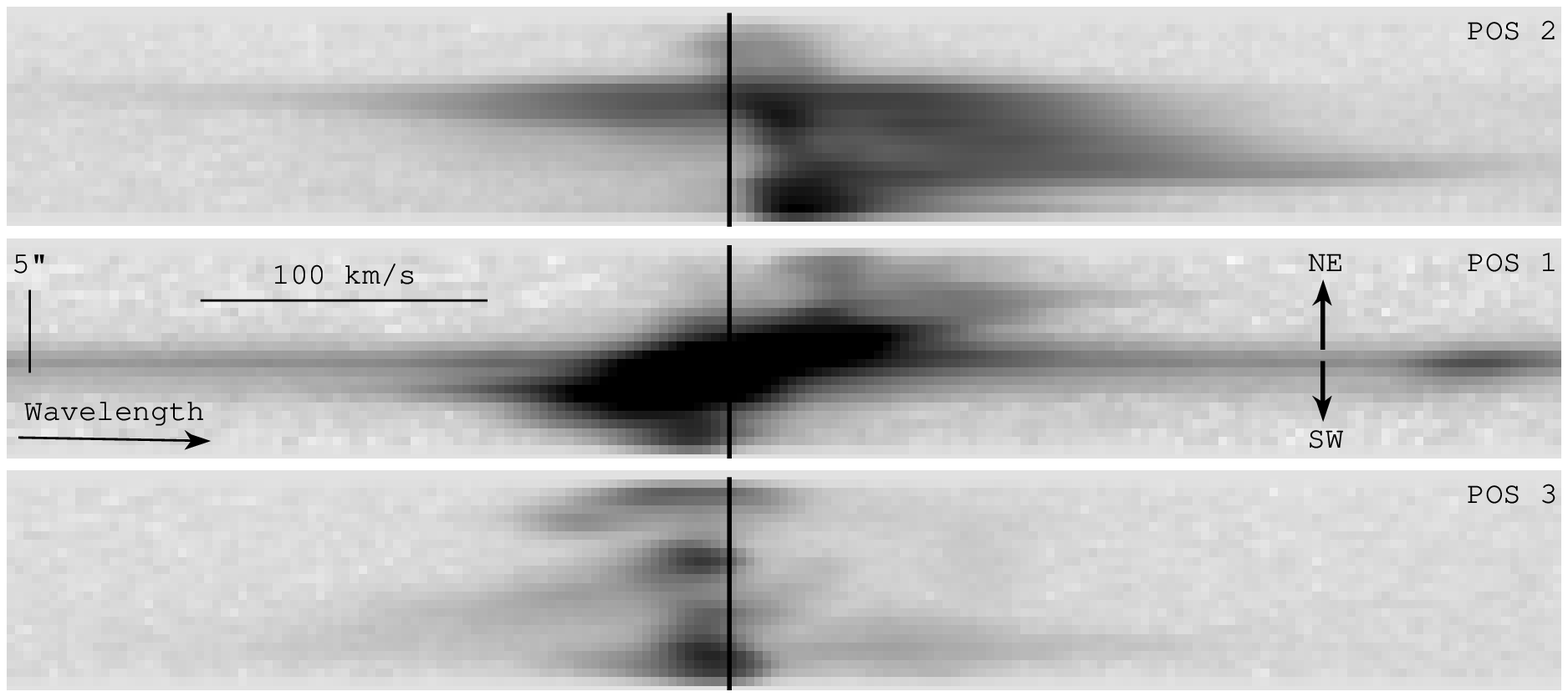}
\includegraphics[width=5.5cm]{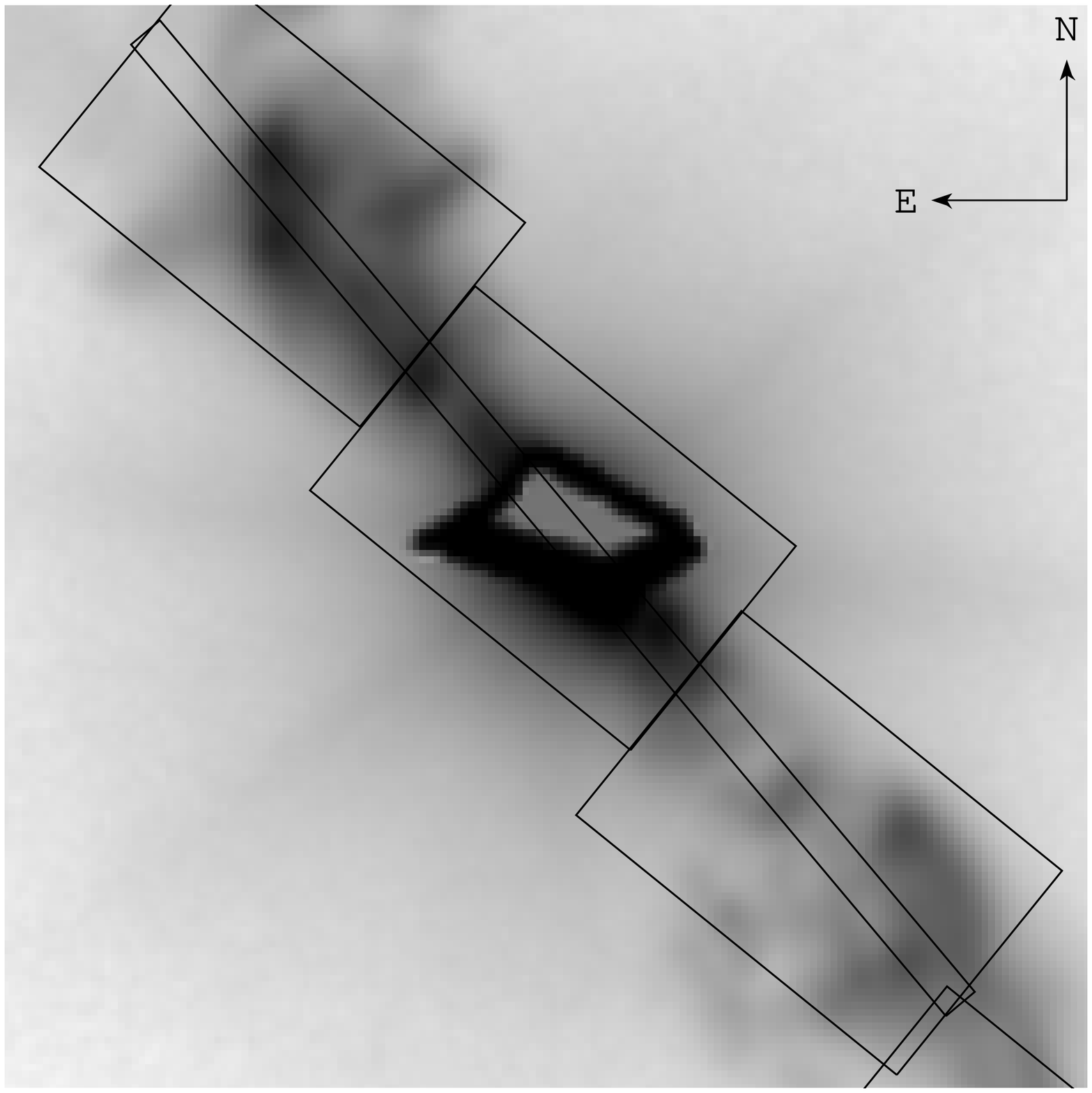}
\caption{\textit{Left.} Extracted long slit spectra with the PA = 40\deg, extending $18''.7$ from the central star in both direction. 
The black vertical line is the 0 \kms radial velocity. In POS1 also the spectrum of the central star is visible. 
\textit{Right.} The extracted long slit shown over the VLT pointings on top of a matched VLT \oiii\ frame. 
The central star is heavily saturated. The FOV is $30''\times30''$, North up, East left, slit width $1''$.
\label{F-ls}}
\end{figure*}

In an attempt to clarify the overall Doppler-shift
kinematics of the jet, and compare with
previous observations, we have extracted from the 3 ARGUS data cubes surrounding the central star a position-velocity 
plot simulating an observation with a long slit,
with a width of 1 arcsec, oriented along the general orientation of the jet  at PA=40$\degr$, and through the central source  (Fig.~\ref{F-ls}).  
\citet{1990ApJ...351L..17H} adopted instead PA=29$\degr$, which was aligned with the inner jet at that time. 
Considering the overall PA change of the jet between the \citet{1990ApJ...351L..17H} and us (about 30 years later)
it clearly indicates that the jet, on large scales, is rotating counterclockwise (CCW). 
The CCW evolution of the jet has also been seen in radio observations \citep{1997ApJ...490..302H}. 

The extracted longslit (Fig.~\ref{F-ls}) confirms the overall structure of the jet.  
However, the figure also highlights the complex variation in velocity profile along the jet.  A simple, ballistically-expanding, 
precessing jet would produce a perfect S-shape, while here we have an overall S-shape but with dramatic variations in the width of the 
S along the slit (much broader in the NE, perhaps reflective of the multiple components like B1 and B2, and with significant 
tails out to very high velocities).  The SW jet, on the other hand, appears more regular in terms of velocity structure but 
much more broken spatially with "gaps" in the emission. 
Furthermore, the artificial long slit spectrum also highlights the acceleration in the inner 
parts of the jet (POS1). This is consistent with the solution proposed in \citet{2004A&A...424..157M} 
that the jet features, after being formed due to increased matter flow at periastron, are accelerated 
inside the first $1''$ and then ejected as bullets.
The outer parts, POS2 and POS3, seem to expand freely. What is unclear throughout the length of the longslit is the contribution of illumination effects - 
perhaps the matter and velocity structure of the two sides of the jet are symmetrical and the observed differences are the result of differing illumination. 
The possible illumination beam visible on the \oiii\ 2007 and 2012 images in NE direction does not seem to move or, at most, moves 
very little. However, the cone is not seen in the southern direction indicating that this may be an important factor in determining the structure of the 
observed position-velocity profile.

%%%%%%%%%%%%%%%%%%%%
%%%%%%%%%%%%%%%%%%%%
\subsection{Comparison with previous works}

When comparing our RV data with previously published data, at a first sight it seems that there is a notable and intriguing difference. 
We find that the NE jet is mostly red-shifted 
and the SW jet mostly blue-shifted, with occasional opposite velocity signs. 
Only in the farthest part of the SW jet covered by our IFU data, a significant redshifted component is detected. 

Previously published data seem to indicate the contradictory behaviour, that is discussed in the following. 
A direct comparison with the RV data of SHKM is difficult because they do not mention if their radial velocities are corrected for the systemic velocity. 
Contrary to our finding, \citet{1985A&A...148..274S} found that the southern jet is mostly red-shifted and the northern one blue-shifted.
Also, they measured radial velocities for features A and B of around -55 \kms\ (based on an \nii\ 6583 \AA\ emission line spectrum), 
which have an opposite sign compared to our measurement for feature B +20 to +70  km/s with respect to the systemic velocity. 
Only near the central star is there some agreement with our measurements, as  \citet{1985A&A...148..274S}  found a negative radial velocity, -20 \kms, in the southern jet.
Overall, \citet{1985A&A...148..274S} say that the southern jet is red-shifted and the northern blue-shifted but if one looks more carefully at their Fig. 5 
it is evident that  \citet{1985A&A...148..274S} have red- and blue-shifted components present at almost all measured positions.

\citet{1990ApJ...351L..17H} present long slit observations of the same emission line as we (\oiii\ 5007~\AA{}) but also find 
that overall the NE jet peak emission is approaching and the SW receding, contrary to our measurements.
Though, it is clear from their Fig. 1 that in the NE up to $7''$ from the centre both blue and redshifted components are equally bright.
Reconstructing their slit on our earlier data we see that most of their SW faint structure is somewhere at our POS4, which is also red-shifted. 
\citet{1999ApJ...522..297H} employed a Fabry-Perot imaging spectrometer to obtain velocity maps of the \nii\ 6583~\AA{} emission line, 
again finding that the northern jet is blue-shifted and the south red-shifted. 
They too remark that the FWHM is extremely large (up to several hundred \kms{}) and that the velocity structure at each 
position is complex, often presenting multiple components, just as we find in our data.

Considering all the above mentioned observations and the nature of our measurements, 
the apparent discrepancies may be explained by brightness variations of the different line profile components,
which would cause that  e.g. a 
previously faint blue component has become significantly brighter than its red component. 
Furthermore, the higher spectral resolution of our data would allow us to better define the 
characteristics of the outflow for different components.

It is unlikely that the differences in radial velocity measurements with previous epochs are mainly 
due to precession effects, given the long $\sim$2000 yr precessing period \citep{1988AJ.....95.1478M}. 

We conclude that, at the intermediate scales of the R Aqr jet investigated in this article,  
most of the observed changes in the morphology and velocity are driven by changes in the illumination and 
ionising conditions of circumstellar gas at the different epochs of observations, as well as of local 
dynamical effects, rather than to a structural change of the entire jet of R Aqr.

Only in the innermost regions, significant structural changes are observed, as shown by \cite{2017A&A...602A..53S}. 
In these regions, some of the discrepancy with 
the RV data between different authors could indeed be related to the actual motion of the ionised material. 
The discrepancies that \cite{2017A&A...602A..53S} 
find gives further indication of the complexity of the R Aqr system.

%%%%%%%%%%%%%%%%%%%%%%%%%%%%%%%%%%%%%%%%%%%
%%%%%%%%%%%%%%%%%%%%%%%%%%%%%%%%%%%%%%%%%%%

\section{Conclusions}\label{S-conc}

We have presented a multi-epoch morphological and kinematical study of the nebula of R Aqr. 

New morphological features (referred to as the arc and loop) outside the known hourglass nebula 
add richness and complexity to the circumbinary 
gas distribution and the mass loss history from the system. 

% 1) Nebula

The large bipolar nebula of R Aqr is expanding ballistically and does not show major structural changes during the observed period. 
An average age of it was calculated to be $T_\mathrm{{bip}}=653\pm35$ years in 1991. Determination of its apparent expansion allows 
us to determine a distance to R~Aqr of 178$\pm$18~pc, consistent, within errors, with \citet{1985A&A...148..274S} and 
\citet{1978ApJ...225..869L}.

% 2) Jet  
The jet is experiencing a more complex evolution. At large scales the jet is mainly expanding radially from the central star. 
However, apparently closer to the central source prominent high velocity lateral movements are detected. 
In addition, structural and brightness changes of several features are detected. 
In the Northern jet, closer to the centre, knotty features tend to 
become elongated until they break into separate blobs. Some features are fading, some brightening, 
and in some cases the position and shape depends on the observed wavelength. 
Even so, the overall S-shape of the jet did not change significantly during the last  30 years. 

Our high resolution radial velocity measurements of the jet present a somewhat controversial behaviour 
with respect to previously published data. We find that the Northern jet is mostly  red-shifted, while the Southern counterpart is blue-shifted. 
Formerly published results show the opposite. We are inclined to believe that this discrepancy is 
due to the general complexity of the line profiles (multiple components, wide wings) and combination of higher spectral 
and spatial resolution of our data, which allows a more detailed view  than previously published long slit spectra. 

The overall conclusion of our study is that the evolution of the jet cannot be described by purely radial expansion, 
and the combined action of changing ionization, illumination, shocks and precession should be added, although it 
is difficult to disentangle the importance of each effect over the others. Continuous monitoring at all wavelengths will help to shed further light into this intriguing object. 

\begin{acknowledgements}

Based on observations made with the Nordic Optical Telescope, operated by the Nordic Optical 
Telescope Scientific Association at the Observatorio del Roque de los Muchachos, La Palma, 
Spain, of the Instituto de Astrofisica de Canarias. The data presented here were obtained with 
ALFOSC, which is provided by the Instituto de Astrofisica de Andalucia (IAA) under a joint 
agreement with the University of Copenhagen and NOTSA.
 Based on observations made with ESO Telescopes at the La Silla Paranal Observatory
under programme ID 089.D-0429 and 090.D-0183.  
% Please notify Eso Library at esolib@eso.org upon acceptance or publication of a  paper based on ESO data, including the bibliographic reference (article title, authors, journal title, volume, year, pages).
This research has made use of the USNOFS Image and Catalogue Archive 
operated by the United States Naval Observatory, Flagstaff Station (http://www.nofs.navy.mil/data/fchpix/).
This work makes use of EURO-VO software TOPCAT. The EURO-VO has been funded by the
European Commission through contracts RI031675 (DCA) and 011892
(VO-TECH) under the 6th Framework Programme and contracts 212104
(AIDA) and 261541 (VO-ICE) under the 7th Framework Programme. 
This research was partially supported by European Social Fund's Doctoral Studies and 
Internationalisation Programme DoRa and Kristjan Jaak Scholarship, 
which are carried out by Foundation Archimedes. 
TL, KV, and IK acknowledge the support of the Estonian Ministry for
Education and Science (grant IUT40-1 and IUT 26-2) and European Regional Development Fund (TK133).
P.A.W acknowledges the support of the French Agence Nationale de la Recherche (ANR), 
under program ANR-12-BS05-0012 "Exo-Atmos".
\end{acknowledgements}

% \begin{thebibliography}{}

\bibliographystyle{aa}
\bibliography{literature}

% \end{thebibliography}

%%%%%%%%%%%%%%%%%%%%%%%%%%%%%%%%%%%%%%%%
%%%%%%%%%%%%%%%%%%%%%%%%%%%%%%%%%%%%%%%%

% \Online

\begin{appendix} %First online appendix

\section{Ballistically expanding jet features.} \label{A-jet}

In this appendix we present more details related to the age analyses of the ballistic jet features. 

\subsection{NE jet}

Due to the structural change in later epochs of feature B$_{\text{SHKM}}$ it is difficult to obtain precise measurements. 
For the proper motion measurements the brightness enhancement was used in filters \ha+\nii, \oii, and \oiii\ . 
%see more from ima_analyse.dat

For the rest of the NE features the magnification method was applied.
For feature C$_{\text{SHKM}}$, the \ha+\nii\ 1991 and 2012 frames needed a further flux correction due to the fact that 
the feature was about twice as bright in 2012 than in 1991, after using the nebula for flux matching.
For rematching, the C$_{\text{SHKM}}$ feature itself was used. Following the same methodology as described in Section \ref{S-neb}, 
the smallest residuals were found in the frame with magnification factor of $M=1.074\pm0.003$. 

% NE4
Due to the change of shape and/or varying exposure time, and hence different level of details detectable in \ha+\nii\ from feature NE4  
between 1991 and 2012, we chose a different filter for the magnification method. After careful visual examination of our data 
we decided to use the \oii\ 2009 and 2012 frames.
Again, the same methodology was followed as for the bipolar nebula, first convolving to the worst seeing 
and then flux matching using the nebula. 
The smallest residuals for NE4 are for $M=1.010\pm0.002$. % Considering that $\Delta t=3.031$ years, an 

 \subsection{SW jet}
 
% loop+S
The longer baseline from 1991 to 2012 in the \ha+\nii\ frames was unsuitable for the magnification method to be used for the feature loop+S 
due to the dramatic change in brightness of knot S as well as the extreme saturation streaks exactly on top of the feature on 1991 frame. 
Given that the nebula is, in general, very similar in the \oii\ filter, we used its 
1997 and 2012 observations (15.14 yrs time interval) to estimate a tentative value for the  magnification of  $M=1.09\pm0.02$. 
%giving an age about $170\pm40$ yrs at 1997.    
% 168.2267953  +-  37.38373228 years 
% see analyses_magnif.dat in /Users/TIINA/ASTRONOOMIA/RAQR/RAQR_1article/RAQR_analyse_magnif/Ha_NII_magnif
% NOTE: can make measurements of the loop edge moving with approximately 0.06 '' per year. See *.reg in RAQR_analyse_magnif/OII_97_12_magnif/loop
L
% SW3
Just as for the NE4 feature, the \oii\ 2009 and 2012 frames were used to derive the age of the SW3 feature via the magnification method. 
The best fitting $M$  was found to be 1.013 $\pm$ 0.002. 
% SW4
In a case of the feature SW4 the same \ha+\nii\ 1991 and 2012 frames, as for the nebular age in 
Section~\ref{S-neb}, were usable. 
% see analyses_magnif.dat in /Users/TIINA/ASTRONOOMIA/RAQR/RAQR_1article/RAQR_analyse_magnif/Ha_NII_magnif

\end{appendix}

\end{document}